\documentclass[aps,pre,reprint,showpacs,superscriptaddress,groupaddress,showkeys,floatfix]{revtex4-1}
\usepackage{srctex}
\usepackage{amsmath}
\usepackage{amsfonts}
\usepackage{amssymb}
\usepackage{epsfig}
\usepackage{graphicx}
\usepackage[usenames,dvipsnames]{xcolor}
\usepackage{soul}
\usepackage{bm}

\begin{document}
\title{Generalized persistence dynamics for active motion}

\author{Francisco J. \surname{Sevilla}}
\email{fjsevilla@fisica.unam.mx}
\thanks{Corresponding author}
\affiliation{Instituto de F\'isica, Universidad Nacional Aut\'onoma de M\'exico,\\
Apdo. Postal 20-364, 01000, Ciudad de M\'exico, M\'exico}

\author{Pavel \surname{Castro-Villarreal}}
\email{pcastrov@unach.mx}
\affiliation{Facultad de Ciencias en F\'isica y Matem\'aticas,
Universidad Aut\'onoma de Chiapas, Carretera Emiliano Zapata, Km. 8, Rancho San Francisco, 29050 Tuxtla Guti\'errez, Chiapas, M\'exico}

\begin{abstract}
We analyze the statistical physics of self-propelled particles from a general theoretical framework that properly describes the most salient characteristic of active motion, \emph{persistence}, in arbitrary spatial dimensions. Such a framework allows the development of a Smoluchowski-like equation for the probability density of finding a particle at a given position and time, without assuming an explicit orientational dynamics of the self-propelling velocity as Langevin-like equation-based models do. Also, the Brownian motion due to thermal fluctuations and the active one due to a general intrinsic persistent motion of the particle are taken into consideration on an equal footing. The persistence of motion is introduced in our formalism in the form of a \emph{two-time memory function}, $K(t,t^{\prime})$. We focus on the consequences when $K(t,t^{\prime})\sim (t/t^{\prime})^{-\eta}\exp[-\Gamma(t-t^{\prime})]$, $\Gamma$ being the characteristic persistence time, and show that it precisely describes a variety of active motion patterns characterized by $\eta$. We find analytical expressions for the experimentally obtainable intermediate scattering function, the time dependence of the mean-squared displacement, and the kurtosis. 
\end{abstract}

\maketitle

\section{Introduction}
The interest in the transport properties of the so-called \emph{active} or \emph{self-propelled} particles has renewed recently due, on the one hand, to their intrinsic out-of-equilibrium nature, 
in clear contrast with the commonly Brownian particles, and, on the other hand, to the designing of artificial particles that use different phoretic mechanisms to self-propel \cite{GompperJPhysCondMatt2020,BechingerRMP2016}. 

A conspicuous feature of active motion is that it is \emph{persistent}, that is to say, the particle retains its direction of motion during a characteristic period of time (called 
\emph{persistence time}) \cite{VicsekPhysRep2012}. However, taking into account the effects of persistence in reduced descriptions of active motion in terms of Smoluchowski-like equations for the marginal probability density that depends only on the particle positions, i.e., when the orientational degrees of freedom that define the direction of motion have been marginalized, has been a difficult task, which we achieve in this paper by the formulation of a general theoretical framework. 

Generally, the reduced descriptions just mentioned lead to the so-called \emph{telegrapher's equation} \cite{GoldsteinQJMAM1951,MasoliverPhysicaA1993,WeissPhysicaA2002}, which emerges as an  approximation valid only in the long-time regime. This has been analyzed for a particular model of active motion in Euclidean spaces in two and three dimensions \cite{SevillaPRE2015,SevillaPRE2016}, and in two-dimensional curved manifolds \cite{CastroPRE2018, ApazaSoftMatter2018J}. For instance, two well-known models that take into account the persistence of active motion have been considered over the years \cite{CatesEPL2013}: the so-called \emph{active Brownian motion} (ABM)\cite{schweitzer2007brownian,KurzthalerSciRep2016} for which the  persistence of orientation is faded by rotational diffusion, and  \emph{run-and-tumble motion} (RTM)\cite{MartensEPJE2012}, for which the persistence of the direction of motion is lost by the instantaneous, temporally uncorrelated tumbling events. In the long-time regime, both frameworks are well approximated by the the telegrapher's equation for the probability density of finding a particle at position $\boldsymbol{x}$ at time $t$, 
$\rho(\boldsymbol{x},t)$, which in $d$ spatial dimensions is given by 
\begin{equation}\label{TelegrapherEq}  \frac{\partial^{2}}{\partial t^{2}}\rho(\boldsymbol{x},t)+ \Gamma \frac{\partial}{\partial t} \rho(\boldsymbol{x},t)=
 c^{2}\nabla^{2}\rho(\boldsymbol{x},t),
\end{equation}
where $\Gamma^{-1}$ is the \emph{persistence time} and $c$ the propagation speed. These two quantities are directly related to (1) the inverse of the rotational diffusion coefficient $D_{R}$ and the particle swimming speed $v_{0}$ in the model of ABM when the identifications $\Gamma=d(d-1)D_{R}$ and $c=v_{0}$ are made, respectively, and (2) the inverse of the tumbling rate $\lambda$ and the particle running speed $v_{0}$ in the model of RTM when $\Gamma=d\lambda$ and $c=v_{0}$ are chosen, respectively. The \emph{persistence length}, defined as $\ell_{p}=c\,\Gamma^{-1}$, characterizes the average length for which the particle maintains the direction of motion. 

As is well known, the telegrapher's equation interpolates between two separated time regimes: the wave equation in the short-time regime and the diffusion equation in the asymptotic one (see Ref. \cite{DunkelPhysRep2009} in the context of relativistic Brownian motion). The good agreement of the results provided by Eq. \eqref{TelegrapherEq} to describe active motion in the long-time regime is intuitively clear and has been discussed before in the case of two-dimensional ABM \cite{SevillaPRE2014}. In this time regime, the faded persistent motion can be effectively described as the motion of Brownian particles diffusing with an effective diffusion coefficient given by $c^{2}/\Gamma$. In contrast to the distortionless propagation of a pulse of active particles in the short-time regime, the telegrapher's equation describes a wave-like motion that suffers from some issues and fails to account for a correct description of active motion, as has been pointed out in Refs. \cite{PorraPRE1997,SevillaPRE2014}. In this paper a generalization of the telegrapher's equation, free of these pathological effects, is derived and analyzed.

A quantity of interest, directly obtained from the probability density of the particle positions and therefore from our theoretical framework, corresponds to the \emph{intermediate scattering function} (ISF). A recent  analysis of the ISF of diffusing Janus particles in two dimensions, concluded that it is possible to discern between ABM and RTM by focusing the analysis in the intermediate-time regime \cite{KurzthalerPRL2018}. It is therefore of interest to establish generalizations of theoretical frameworks whose validity comprises the whole time regimes to describe active motion, particularly at the intermediate and at the short ones. 

In this paper we propose a general theoretical framework to analyze a {\it generalized active Brownian motion} (GABM) model that incorporates the rather important effects of the persistence of active motion through a \emph{memory function}. This last one models the pattern of motion induced by the particular internal mechanism of the particle self-propulsion (notice that this formulation differs of the one considered in Refs. \cite{SevillaChapter2018,SevillaPRE2019b}, where the internal dynamics of the active particle is embedded in the memory function of the generalized Langevin equation). Here we consider a memory function whose time dependence takes into account the persistence of motion by modifying, in the short-time regime only, the exponentially decaying memory function, $e^{-\Gamma t}$, that appears in the telegrapher's equation \eqref{TelegrapherEq} when this is rewritten as \cite{ForsterBook}
\begin{equation}
\frac{\partial}{\partial t}\rho(\boldsymbol{x},t)=c^{2}\int_{0}^{t}ds\, e^{-\Gamma(t-s)}\, \nabla^{2}\rho(\boldsymbol{x},s).
\end{equation}
We show afterwards that by modifying the dynamics in this respect, not only is the poor behavior of the telegrapher-equation solutions in high dimensions corrected, but also reveals well-behaved wavelike patterns of motion.

Furthermore, our analysis considers the important influence of the fluctuations exerted by the surroundings which give rise to Brownian motion, characterized by the translational diffusion coefficient $D_\text{T}$. To our knowledge, the role of thermal fluctuations in the telegrapher equation has been treated only very recently in Ref. \cite{DulaneyPRE2020}, as an approximation to the wave propagation of active Brownian particles. All these elements together allow us to carry out  an explicit comparison between the \emph{intermediate scattering function} of the model introduced here and the corresponding one for those well-known models of persistent motion, namely, active Brownian motion and run-and-tumble motion. Additionally we obtain the explicit time dependence of the mean-squared displacement (MSD) $\langle\boldsymbol{x}^{2}(t)\rangle$ and of the kurtosis $\kappa(t)$ for the active component of motion. 

Our paper is organized as follows: In Sec. \ref{Sect-II} the theoretical framework to study the GABM is introduced, and within this, a generalized telegraphers equation for active motion is presented and analyzed. We provide explicit analytical expressions for the time dependence of the MSD and the kurtosis related to the active part of motion. In Sec. \ref{Sect-III}, we give a comparison between the GABM and the theoretical predictions made by the ABM and RTM models. Finally, in Sec. \ref{Sec. IV} we give our concluding remarks and perspectives. 

\section{\label{Sect-II} Generalized Active Brownian Motion}

In this section we provide a probabilistic description of the stochastic persistent motion of a single active particle under the influence of thermal noise. Such a description is adequate in the dilute regime of active systems, i.e., far from the crowded regime where \emph{motility induced phase separation} is observed. The main goal is to present a theoretical framework to analyze active motion in the presence of thermal noise, from the knowledge of the statistical properties of active motion in the absence of it. This is carried out in terms of a transport equation based on the continuity equation 
\begin{equation}
\frac{\partial}{\partial t}P(\boldsymbol{x}, t)+\nabla\cdot\boldsymbol{J}(\boldsymbol{x},t)=0,
\label{continuity}
\end{equation}
which endows the conservation of probability by coupling the change in time of the probability density function (PDF) $P(\boldsymbol{x},t)$ of finding 
a particle at the position $\boldsymbol{x}$ at the time $t$, with the probability current $\boldsymbol{J}(\boldsymbol{x},t)$, which indicates the 
change of probability per unit of area per unit of time.

Our departure point relies on the assumption that the statistical properties of active motion, which arises from the particle's internal mechanisms of self-propulsion, are known and encompassed by the probability density $P_{a}(\boldsymbol{x},t)$ and the probability current $\boldsymbol{J}_{\text{a}}(\boldsymbol{x},t)$, which satisfy the continuity equation 
\begin{equation}
\frac{\partial}{\partial t}P_{a}(\boldsymbol{x}, t)+\nabla\cdot\boldsymbol{J}_{\text{a}}(\boldsymbol{x},t)=0.
\label{continuity-active}
\end{equation}
Equations \eqref{continuity} and \eqref{continuity-active} are incomplete until ``constitutive relations'' between the probability currents $J(\boldsymbol{x},t)$, $J_\text{a}(\boldsymbol{x},t)$, and the probability densities $P(\boldsymbol{x},t)$ and $P_\text{a}(\boldsymbol{x},t)$, are provided. We show afterwards, once the constitutive relations are introduced, that Eq. \eqref{continuity} reduces to Eq. \eqref{continuity-active} in the absence of thermal noise. 
A first "constitutive relation" is given by 
\begin{multline}\label{TotalCurrent}
\boldsymbol{J}(\boldsymbol{x},t)=-D_{\text{T}}\nabla P(\boldsymbol{x},t)+
\int d\boldsymbol{x}^{\prime}\,G(\boldsymbol{x}-\boldsymbol{x}^{\prime},t)\boldsymbol{J}_{\text{a}}(\boldsymbol{x}^{\prime},t),
\end{multline}
leaving the second ``constitutive relation'' to be specified later [see Eq. \eqref{ActiveCurrent} below]. Besides the constitutive relation \eqref{TotalCurrent}, we need also to specify the relation between $P(\boldsymbol{x},t)$ and $P_\text{a}(\boldsymbol{x},t)$, which is given as the convolution of the probability density of active motion $P_\text{a}(\boldsymbol{x},t)$ with the Gaussian propagator $G(x,t)$:
\begin{equation}\label{TotalPDF}
P(\boldsymbol{x},t)=\int d\boldsymbol{x}^{\prime}\, G(\boldsymbol{x}-\boldsymbol{x}^{\prime},t)P_{\text{a}}(\boldsymbol{x}^{\prime},t).
\end{equation}

In Eqs. \eqref{TotalCurrent} and \eqref{TotalPDF}, $G(\boldsymbol{x}-\boldsymbol{x}^{\prime},t)$ is the $d$-dimensional Gaussian distribution, also called the \emph{Brownian propagator} solution of the diffusion equation $\partial_{t}G(\boldsymbol{x},t)=D_\text{T}\nabla^{2}G(\boldsymbol{x},t)$, and it is given by $\exp\bigl\{-(\boldsymbol{x}-\boldsymbol{x}^{\prime})^{2}/4D_{\text{T}}t\bigr\}/(4\pi D_{\text{T}}t)^{d/2}$ for the initial condition $G(\boldsymbol{x},0)=\delta(\boldsymbol{x})$. This describes the fluctuating motion due to the effects of an external source of stochastic motion, like the one provided by a thermal bath which leads to \emph{passive} motion usually referred to as Brownian motion, frequently not negligible for microorganisms or artificial particles in the range of a few micrometers. It is now easy to show that in the absence of thermal noise, i.e., $D_\text{T}\rightarrow0$ and thus $G(\boldsymbol{x}-\boldsymbol{x}^{\prime},t)\rightarrow\delta(\boldsymbol{x}-\boldsymbol{x}^{\prime})$, we have that $\boldsymbol{J}(\boldsymbol{x},t)=\boldsymbol{J}_\text{a}(\boldsymbol{x},t)$ in Eq. \eqref{TotalCurrent} and $P(\boldsymbol{x},t)=P_\text{a}(\boldsymbol{x},t)$ in Eq. \eqref{TotalPDF}.

We turn now to discuss the meaning of the total probability current $\boldsymbol{J}(\boldsymbol{x},t)$ in Eq. \eqref{TotalCurrent} and of the total probability $P(\boldsymbol{x},t)$ in Eq. \eqref{TotalPDF}. Consider first the former one. The first term in the right-hand side of Eq. \eqref{TotalCurrent} corresponds to the standard Fick's law with $D_{\text{T}}=k_{B}T/\gamma$ being the translational diffusion coefficient that characterizes the influence of the surroundings at uniform temperature $T$, $k_{B}$ the Boltzmann constant, and $\gamma$ the friction coefficient that results of the interaction between the particle and the external source of heat, a fluid in most cases. The second term takes into account the contribution of the current due to active motion, written as the convolution of $G(\boldsymbol{x}-\boldsymbol{x}^{\prime},t)$ and the current of active motion $\boldsymbol{J}_{\text{a}}(\boldsymbol{x},t)$. The physics of the second term in the right-hand side of Eq. \eqref{TotalCurrent} can be understood under the following rationale. The internal dynamics of self-propulsion characterized by the active current $\boldsymbol{J}_{a}(\boldsymbol{x},t)$, gives rise to the advection term $\nabla\cdot\big[\boldsymbol{V}_{\text{a}}(\boldsymbol{x},t)P(\boldsymbol{x},t)\big]$ in the 
transport equation \eqref{continuity},
\begin{equation}
 \frac{\partial}{\partial t}P(\boldsymbol{x}, t)+\nabla\cdot\big[\boldsymbol{V}_\text{a}(\boldsymbol{x},t)P(\boldsymbol{x},t)\big] 
=D_{\text{T}}\nabla^{2}P(\boldsymbol{x},t),
\end{equation}
where the probability current $\boldsymbol{V}_\text{a}(\boldsymbol{x},t)P(\boldsymbol{x},t)\equiv\int 
d\boldsymbol{x}^{\prime}G(\boldsymbol{x}-\boldsymbol{x}^{\prime},t)\boldsymbol{J}_{\text{a}}(\boldsymbol{x}^{\prime},t)$ defines the active velocity 
field $\boldsymbol{V}_\text{a}(\boldsymbol{x},t)$, as can be verified by substituting \eqref{TotalCurrent} in \eqref{continuity}. Locally, this 
probability current is the result of the contributions of the active current $\boldsymbol{J}_{a}(\boldsymbol{x}^{\prime},t)$ at each point in 
space, weighted by the Gaussian propagator.

Second, Eq. \eqref{TotalPDF} simply expresses that the change of the particle position in the time interval $\Delta t$, $\Delta\boldsymbol{x}(t;\Delta t)\equiv\boldsymbol{x}(t+\Delta)-\boldsymbol{x}(t)$, is decomposed as the sum of a random displacement due to active motion $\Delta\boldsymbol{x}_\text{a}(t;\Delta t)\equiv\int_{t}^{t+\Delta t}ds\, \boldsymbol{v}_\text{a}(s)$,  and another due to Brownian motion $\Delta\boldsymbol{x}_\text{B}(t;\Delta t)\equiv\int_{t}^{t+\Delta t}ds\, \boldsymbol{\xi}_{\boldsymbol{B}}(s)$. Here $\boldsymbol{v}_\text{a}(t)$ denotes the stochastic swimming velocity of the particle and $\boldsymbol{\xi}_{\boldsymbol{B}}(t)$ is three-dimensional Gaussian-white noise modeling the thermal fluctuations. This description can be embodied in the stochastic differential equation
\begin{equation}
    \frac{d}{dt}\boldsymbol{x}(t)=\boldsymbol{v}_\text{a}(t)+\boldsymbol{\xi}_{\boldsymbol{B}}(t).
\end{equation}
In addition, one  can prove that both Eqs. (\ref{TotalCurrent}) and (\ref{TotalPDF}) are self-consistent with the active continuity equation (\ref{continuity-active}) (see Appendix \ref{sect:Proof}).

In order to determine the statistical properties of the persistent part of motion, we require to make explicit the second constitutive relation between the current $\boldsymbol{J}_{\text{a}}(\boldsymbol{x},t)$ and the probability density $P_{\text{a}}(\boldsymbol{x},t)$ in Eq. \eqref{continuity-active}. 
Here, we 
assume the generalized form of the Fick's law,
\begin{equation}\label{ActiveCurrent}
\boldsymbol{J}_{\text{a}}(\boldsymbol{x},t)=-\nabla\int_{0}^{t}dt^{\prime}K(t,t^{\prime}) P_{\text{a}}(\boldsymbol{x},t^{\prime}),
\end{equation}
where $K(t,t^{\prime})$ denotes a memory function that embeds the implicit dynamics of the particle swimming direction and thus, the patterns of active motion. This constitutive relation leads to a generalization of the  diffusion equation for $P_\text{a}(\boldsymbol{x},t)$ when $\boldsymbol{J}_\text{a}(\boldsymbol{x},t)$ is substituted in Eq. \eqref{continuity-active}. Such a generalization considers a temporal nonlocality, and to the knowledge of the authors, no simple stochastic process in the form of a stochastic differential equation for $\boldsymbol{v}_\text{a}(t)$ gives rise to it (the analysis of this aspect alone deserves attention an will be discussed elsewhere). In some cases of physical interest, it can be considered invariant under time translations, i.e., $K(t,t^{\prime})=K(t-t^{\prime})$; however in general, such memory can be nonlocal \cite{SevillaPRE2020} in space too, but such a complication is unnecessary at the 
level of description of the present paper. We want to point out in advance that the two first even moments of $P(\boldsymbol{x},t)$ can be written in terms of the memory 
function $K(t,t^{\prime})$ \cite{KenkreSevilla2007} as
\begin{equation}
\langle \boldsymbol{x}^2(t)\rangle=2dD_{T}t+2d\int_{0}^{t}dt^{\prime}\int_{0}^{t^{\prime}}ds K(t^{\prime}, s),
\label{generalsolMSD}
\end{equation}
and
\begin{multline}
\langle \boldsymbol{x}^4(t)\rangle=4(d+2)\Biggl[D_{T}\int_{0}^{t}ds\langle \boldsymbol{x}^2(s)\rangle\\+\int_{0}^{t}dt^{\prime}\int_{0}^{t^{\prime}}ds 
K(t^{\prime}, s)\langle \boldsymbol{x}^2(s)\rangle\Biggr],
\label{generalsol4moment}
\end{multline}
which reveal, explicitly, the role of the memory function.

The general scope of our theoretical approach, endowed in the memory function $K(t,t^{\prime})$, can be reviewed as follows. Lets first recall that active Brownian and run-and-tumble motion provide an explicit stochastic dynamics of self-propulsion or, equivalently, of the pattern of active motion, which in the long-time regime, the details of the fluctuating dynamics of the swimming direction for both models can be cast into the zero-ranged memory function $K(t)=D_{\text{eff}}\, \delta(t)$, with $D_{\text{eff}}$ an effective diffusion coefficient which is identified with $v_{0}^{2}/\Gamma$, $v_{0}$ the particle swimming speed and $\Gamma^{-1}$ the persistence time. A first correction that incorporates the effects 
of persistence on both models is given by the exponentially damped memory function $K(t)=D_{\text{eff}}\Gamma\, e^{-\Gamma t}$, which is valid in the long-time 
regime and leads to the telegrapher's equation \eqref{TelegrapherEq} after substitution of $K(t)$ in Eq. \eqref{continuity-active} with $\rho(\boldsymbol{x},t)=P_{\text{a}}(\boldsymbol{x},t)$ \cite{ForsterBook,DunkelPhysRep2009}. Such a memory function can be rigorously obtained from generic models of active motion (ABM, RTM) in the long-time regime, where the polar approximation is valid, i.e., when one can neglect the multipoles of order higher than one in the multipolar expansion of the complete probability density distribution \cite{CatesEPL2013,CastroPRE2018,SevillaPRE2020} (see Appendix \ref{AppendixActiveMotion}). Thus, proper generalizations of the memory function $K(t,t^{\prime})$, would lead to physically meaning descriptions of different patterns of active motion, including ABM and RTM.

Finally, we want to point out that although the theoretical framework posed in this paper is limited to the description of the free motion of an active particle, it can be extended to different situations with almost no change. For instance, when the active particle moves on a curved surface we simply replace the vector derivative $\nabla$ with the covariant derivative compatible with the metric tensor of the surface. This situation in particular, is of interest for the study of the combined effects of persistent motion and the surface curvature on the active transport processes such as the ones studied in Ref. \cite{CastroPRE2018}. Another case of interest corresponds to the inclusion of the effects of confinement due to an external field; in this case we can analyze the role of thermal fluctuations if the statistical properties of active motion under the influence of the external field in the absence of thermal noise are known. Simply, the term $\boldsymbol{J}_\text{ext}(\boldsymbol{x},t)\propto P_\text{a}(\boldsymbol{x}, t)\nabla \varphi(\boldsymbol{x})$, must be added to the probability current $\boldsymbol{J}_\text{a}(\boldsymbol{x},t)$ \eqref{ActiveCurrent}, without further changes in Eqs. \eqref{continuity}, \eqref{continuity-active}, and \eqref{TotalPDF}, where $\varphi(\boldsymbol{x})$ is the external potential field acting on the active particles \cite{chaikin_lubensky_1995}. The case when active motion is confined by ``hard-wall'' boundaries, as occurs in many experimental situations, can be treated in a similar fashion, however; the restrictions of zero flux at the boundaries must be taken into account as much as for the propagator of thermal diffusion (this can be expanded in eigenfunctions of the diffusion equation that satisfies the boundary conditions), as for the solution of active motion in the absence of thermal noise. These extensions are of interest, and a discussion of the possible solutions will be presented elsewhere.

\subsection{A Persistent dynamics through a memory function}
In this paper we argue that to fulfill the implicit dynamics of the swimming vector of active particles, not only in the long-time regime but also in the whole span of time, a power-law dependence should be incorporated into the memory function $K(t,t^{\prime})$. We propose, instead of the time-translational invariant 
memory function in \eqref{ActiveCurrent}, an exponentially damped power-law memory function given explicitly by
\begin{equation}\label{ActiveKernel}
K(t,t^{\prime})=c^{2}\frac{t^{-\eta}e^{-\Gamma t}}{{t^{\prime}}^{-\eta}e^{-\Gamma t^{\prime}}},\quad t\le t^{\prime},
\end{equation}
where $\eta$ is a non-negative dimensionless parameter that takes into account how fast the ratio $t/t^{\prime}$ diminishes for $t\gg t^{\prime}$ (or grows as $t\ll t^{\prime}$), $c$ is another parameter with units of speed, and $\Gamma^{-1}$ is the persistence time. As is made clear later in this paper, $\eta$ defines an effective ``temporal dimension'' whose value is intimately related to the particle's effective speed in the short-time regime. Though \eqref{ActiveKernel} is not invariant under time translations, the consequences of such a particular form are 
relevant for the description of active motion as is discussed in the following. Notice that for $\eta=0$ we recover the exponential memory function 
that leads to the telegrapher's equation \eqref{TelegrapherEq}.

After substitution of the memory function \eqref{ActiveKernel} in Eq. \eqref{continuity-active}, we get a closed equation for the probability density $P_{\text{a}}(\boldsymbol{x},t)$:
\begin{eqnarray}\label{GeneralizedEq}
 \left[\frac{\partial^{2}}{\partial t^{2}}+ \left(\frac{\eta}{t}+\Gamma\right)\frac{\partial}{\partial t}\right]P_{\text{a}}(\boldsymbol{x},t)=
 c^{2}\nabla^2 P_{\text{a}}(\boldsymbol{x},t).
\end{eqnarray}
The second term in the left-hand side of the last expression considers two effects: (1) the short-time effects of active motion indicated by $\eta$; and (2) the dissipation as occurs in the telegrapher's equation \eqref{TelegrapherEq}, with $\Gamma^{-1}$ the timescale that characterizes the correlation time of the orientation of the particle direction of motion. Initial conditions corresponding to a pulse at the origin of coordinates that propagates with vanishing initial flux are of interest and are denoted by $P_{\text{a}}(\boldsymbol{x},0)=\delta(\boldsymbol{x})$, and $\partial P_{\text{a}}(\boldsymbol{x},0)/\partial t=0$. We want to point out in passing that in the dissipationless case, i.e., $\Gamma=0$, Eq. \eqref{GeneralizedEq} coincides with the generalization of the wave equation of Bietenholz and Giambiagi \cite{BietenholzJMathP1995} in $d$ isotropic spatial dimensions and in $d_{\eta}=\eta+1$ isotropic ``temporal dimensions.'' 

We show that the solutions to Eq. \eqref{GeneralizedEq} describe active motion in the whole time regime. Furthermore, since such a solution gives directly $P(\boldsymbol{x},t)$, the probability density of finding a particle at $\boldsymbol{x}$ at time $t$, independent of the direction of motion motion, our theoretical framework dispenses with the standard approach of getting a hierarchy of coupled equations for the moments of the complete distribution and truncating the hierarchy at some suitable order.

\subsection{Probability density function for the effective active motion $P_{\text{a}}(\boldsymbol{x},t)$}
\begin{figure*}[t]
\includegraphics[trim=15 35 10 20,width=\textwidth]{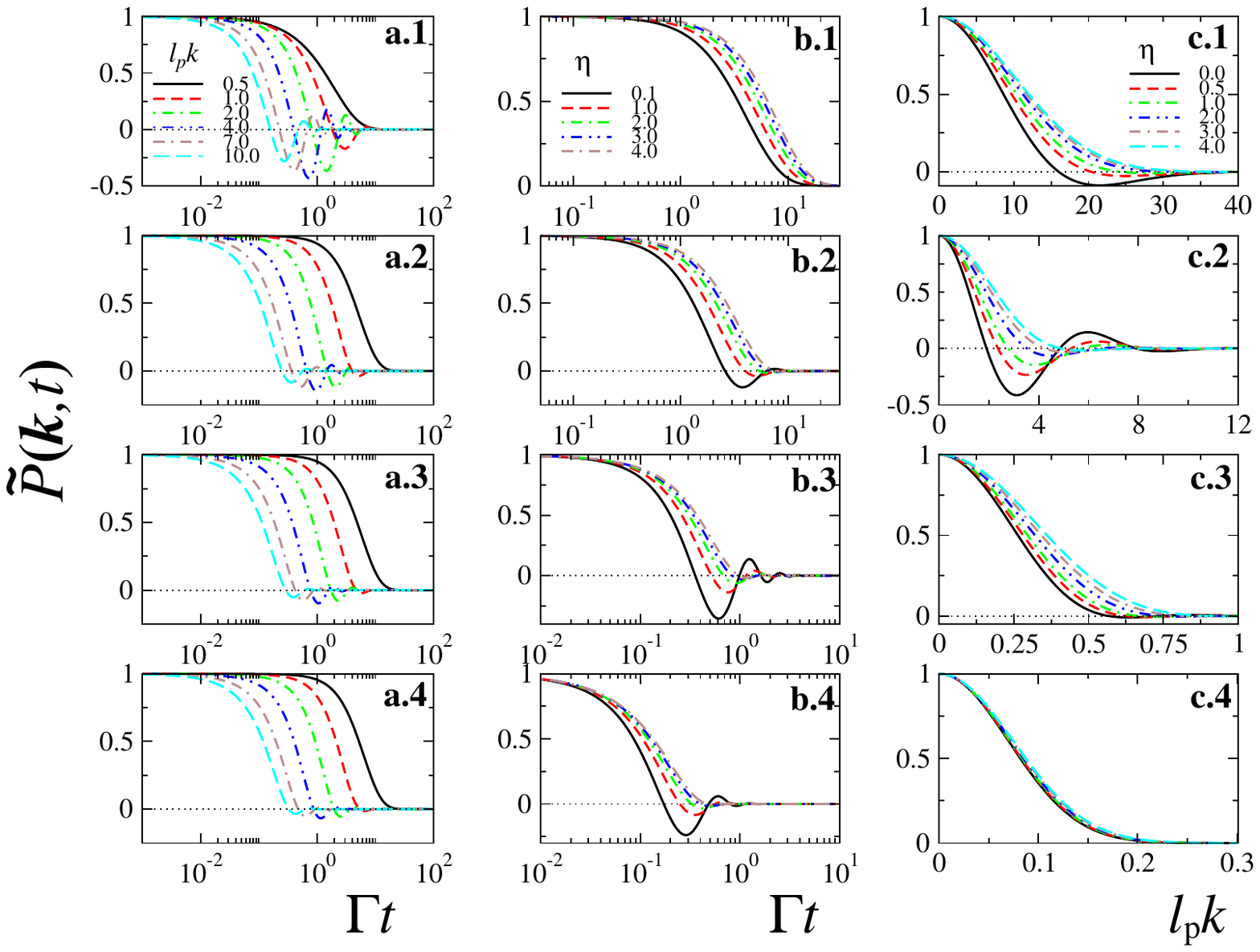}
\caption{(Color online) Intermediate scattering function $\widetilde{P}(\boldsymbol{k},t)$ as given in \eqref{TotalPDF} with an inverse P\'eclet 
number $\widetilde{D}=D_{\text{T}}/\Gamma \ell_{\text{p}}^{2}=0.038$ obtained from the experimental data in Ref. \cite{KurzthalerPRL2018}. Columns {\bf a} and {\bf b} show $\widetilde{P}(\boldsymbol{k},t)$ as function of the dimensionless time $\Gamma t$ , and column {\bf c} as function of the dimensionless wave vector $\ell_{\text{p}}k$. The panels in column {\bf a} correspond to $\eta=0$ ({\bf a.1}), $\eta=1$ ({\bf a.2}), $\eta=3/2$ ({\bf a.3})  and $\eta=2$ ({\bf a.4}), where $\widetilde{P}(\boldsymbol{k},t)$ is shown for different values of 
the dimensionless wave vector $\ell_{\text{p}}k$ 0.5 (solid black line), 1.0 (dashed red line), 2.0 (dash-dotted green line), 4.0 (dash-double-dotted blue line), 7.0 (double-dash-dotted brown line), and 10.0 (long-dashed cyan line). The panels in column {\bf b} show $\widetilde{P}(\boldsymbol{k},t)$ for: $\ell_{\text{p}}k=0.5$ ({\bf b.1}), 
$\ell_{\text{p}}k=1$ ({\bf b.2}), $\ell_{\text{p}}k=5$ ({\bf b.3}), and $\ell_{\text{p}}k=10$ ({\bf b.4}); different curves correspond to $\eta=0.1$ (solid black line), 1.0 (dashed red line), 2.0 (dash-dotted green line), 3.0 (dash-double-dotted blue line) and 4.0 (long-dashed brown line). Column {\bf c} shows $\widetilde{P}(\boldsymbol{k},t)$ as a function of the dimensionless wave vector $\ell_{\text{p}}k$ at different times: $\Gamma t=0.1$ 
({\bf c.1}); $\Gamma t=1.0$ ({\bf c.2}); $\Gamma t=10.0$ ({\bf c.3}); $\Gamma t=100.0$ ({\bf c.4}) for different values of $\eta$: 0 (solid black line), 0.5 (dashed red line), 1.0 (dash-dotted green line), 2.0 (dash-double-dotted blue line), 3.0 (double-dash-dotted brown line), and 4.0 (long-dashed cyan line).}
\label{PDFk_t}
\end{figure*}
We can take advantage of the simple physical circumstances considered here, namely, that active particles diffuse freely and isotropically in open space, in order to find solutions to Eq. \eqref{GeneralizedEq} in Fourier space. After Fourier transforming Eq. \eqref{GeneralizedEq} we get
\begin{equation}\label{TransformedGeneralizedEqFourier}
 \frac{d^{2}}{dt^{2}}\widetilde{P}_{\text{a}}(\boldsymbol{k},t)+\left(\frac{\eta}{t}+\Gamma\right)\frac{d}{dt}\widetilde{P}_{\text{a}}(\boldsymbol{k},t)+c^{2}\boldsymbol{k}^{2}\widetilde{P}_{\text{a}}(\boldsymbol{k},t)=0,
\end{equation}
where we have chosen
\begin{equation}
\widetilde{P}_{\text{a}}(\boldsymbol{k},t)=\int d\boldsymbol{x}\, e^{-i\boldsymbol{x}\cdot \boldsymbol{k}}P_{\text{a}}(\boldsymbol{x},t)
\end{equation}
as the direct Fourier transform, and $\boldsymbol{k}$ denotes the two-dimensional Fourier variable (wave vector) conjugate to $\boldsymbol{x}$. As usual, the inverse transform involves the kernel $(2\pi)^{-d}e^{i\boldsymbol{k}\cdot\boldsymbol{x}}$. Equation \eqref{TransformedGeneralizedEqFourier} is complemented by the corresponding initial conditions $\widetilde{P}_{\text{a}}(\boldsymbol{k},0)=1$ and $\left(\partial/\partial t\right)\widetilde{P}_{\text{a}}(\boldsymbol{k},0)=0$.

The rotational symmetry of Eq. \eqref{TransformedGeneralizedEqFourier} implies that its solution depends only on $k$; the magnitude of $\boldsymbol{k}$, then $\widetilde{P}_{\text{a}}(\boldsymbol{k},t)=\widetilde{P}_{\text{a}}(k,t)$ and for $\eta>0$ the solution can be obtained by solving a Kummer differential equation as it is shown in Appendix \ref{appDiff}. Consequently, the solution is written in terms of the Kummer function $_{1}F_{1}(a,b;z)=\sum_{\ell=0}^{\infty}\frac{(a)_{\ell}}{(b)_{\ell}}\frac{z^{\ell}}{\ell!}$; in particular, for $\eta=0$ it can be simplified using explicit expressions of this  function  \cite{Abramowitz1964},
\begin{widetext}
\begin{equation}\label{Solution}
    \widetilde{P}_{\text{a}}(\boldsymbol{k},t)=e^{-\Gamma t/2}\left\{\begin{array}{rc}
    \cosh\left(\sigma_{k}\frac{\Gamma t}{2}\right)+\frac{1}{\sigma_{k}}\sinh\left(\sigma_{k}\frac{\Gamma t}{2}\right), & \eta=0\\
    e^{-\sigma_{k}\Gamma t/2}\, _{1}F_{1}\left[\frac{\eta}{2}\left(1+\frac{1}{\sigma_{k}}\right),\eta;\sigma_{k}\, \Gamma t\right], & \eta>0\\
    \end{array}
    \right.,
\end{equation}
\end{widetext}
where $\sigma_{k}=\sqrt{1-4\ell_{\text{p}}^{2}k^{2}}$ is a dimensionless, rotationally invariant function of the dimensionless quantity $\ell_{p}k$.

Thus, for $\eta>0$ the complete probability distribution \eqref{TotalPDF}, or ISF \cite{KurzthalerSciRep2016}, is a function of the magnitude of $\boldsymbol{k}$ and is given by 
\begin{multline}\label{TotalPDF-Fourier}
    \widetilde{P}(\boldsymbol{k},t)=\exp\left\{- t\left[D_{\text{T}}k^{2}+\frac{\Gamma}{2}(1+\sigma_{k})\right]\right\}\times\\
    _{1}F_{1}\left[\frac{\eta}{2}\left(1+\frac{1}{\sigma_{k}}\right),\eta;\sigma_{k}\, \Gamma t\right],
\end{multline}
where the factor $\exp\{-D_{\text{T}}k^{2}t\}$ is the $d$-dimensional Fourier transform of the Gaussian propagator $G(\boldsymbol{x},t)$ 
that appears in \eqref{TotalPDF}. The appearance of the dimensionless parameter $\eta$ and the inverse P\'eclet number $\widetilde{D}_{\text{T}}=D_{\text{T}}/(\Gamma \ell_{\text{p}}^{2})$ is made explicit in the solution \eqref{TotalPDF-Fourier} when this is written as function of the dimensionless wave vector $\ell_{p}k$. As is evident from the expression \eqref{TotalPDF-Fourier}, the behavior is independent of the spatial dimensionality of the system since the dependence is on the wave-vector magnitude only.

The ISF is shown in Fig. \ref{PDFk_t} for $\eta\ge0$ and for the reference value of the inverse P\'eclet number  $\widetilde{D}_{T}=0.038$ (computed from the data obtained from Janus particle trajectories moving in two dimensions in Ref. \cite{KurzthalerPRL2018}), as a function of the following:
\begin{itemize}
\item[({\bf a})] The dimensionless time, $\Gamma t$, for different values of $l_{p}k$ marked by different line style. The effects of the parameter $\eta$ are shown in the different panels of column {\bf a}:  ({\bf a.1}) $\eta=0$, ({\bf a.2}) $\eta=1$, ({\bf a.3}) $\eta=3/2$ and ({\bf a.4}) $\eta=2$. This behavior is qualitatively similar to the ISF obtained experimentally for Janus particles diffusing in two dimensions \cite{KurzthalerPRL2018}.
\item[({\bf b})] The same variable, $\Gamma t$, for different values of $\eta$ marked by different line style, at different values of 
$\ell_{\text{p}}k$ shown in the different panels of column {\bf b}, that is, {\bf (b.1)} $\ell_{\text{p}}k=0.5$, ({\bf b.2}) $\ell_{\text{p}}k=1$, 
({\bf b.3}) $\ell_{\text{p}}k=5$, and $\ell_{\text{p}}k=10$ ({\bf b.4}) .
    \item[{\bf (c)}] The dimensionless magnitude of the wave vector, $\ell_{\text{p}}k$, at the dimensionless time $\Gamma t=0.1 ({\bf c.1})$ and $1.0$ 
({\bf c.2}) for which the characteristic oscillations of active motion are shown, and $\Gamma t=10$ ({\bf c.3}) and $\Gamma t=100$ ({\bf c.4}), for which 
the oscillations are damped out and the solutions converge to a universal Gaussian distribution independently of $\eta$.
\end{itemize}

In the short-time regime, $\Gamma t\ll1$, the damping term (the one proportional to $\Gamma$) can be neglected in Eq. \eqref{TransformedGeneralizedEqFourier}. In this regime the active probability density for the initial data chosen is given by
\begin{equation}\label{PDFactive-S-T}
    \widetilde{P}_{\text{a,s-t}}(\boldsymbol{k},t)\simeq\Gamma\left(d_{\eta}/2\right)\frac{J_{d_{\eta}/2-1}(kct)}{(kct/2)^{d_{\eta}/2-1}},
\end{equation}
where $J_{\alpha}(z)$ is the $\alpha$-th order Bessel function of the first kind, $\Gamma(z)$ is the standard Gamma function, and we have explicitly shown the dependence on the ``temporal 
dimension'' $d_{\eta}$ introduced by Bietenholz and Giambiagi \cite{BietenholzJMathP1995}. Remarkably, it can 
be shown that Eq. \eqref{PDFactive-S-T} coincides exactly with the intermediate scattering function obtained in the short-time regime, from the expressions of the ABM given in Ref. \cite{KurzthalerSciRep2016} and from the RTM given in Ref. \cite{KurzthalerPRL2018} when $d_{\eta}=d$. The general expression \eqref{PDFactive-S-T} gives rise to a whole family of rotationally symmetric pulses that propagates with speed $c$:
\begin{multline}\label{Pulses}
     P_{\text{a,s-t}}(\boldsymbol{x},t)=\frac{\Gamma(d_\eta/2)}{2^{d-1}\pi^{d/2}}\int_{0}^{\infty}dk\, k^{d-1}\times\\
     \frac{J_{d_{\eta}/2-1}(kct)}{(kct/2)^{d_{\eta}/2-1}}
     \frac{J_{d/2-1}(kx)}{(kx/2)^{d/2-1}},
\end{multline}
where it can be noticed that the spatial dimensionality $d$, and the temporal one $d_{\eta}$, play a symmetric role. In particular, it is noticeable that Eq. \eqref{Pulses} reveals a connection between $\eta$ and the spatial dimensionality of the system $d$, for if $d_\eta$ is chosen to be equal to the spatial dimension $d$, we get that the probability density \eqref{PDFactive-S-T} can be written in the spatial coordinates as the rotationally symmetric sharp pulse:
\begin{equation}\label{PDFactive-Pulse}
 P_{\text{a,s-t}}(\boldsymbol{x},t)=\frac{\delta(x-ct)}{\Omega_{d}x^{d-1}},
\end{equation}
which propagates without distortion in any dimension and is free from some of the undesirable features of the propagating pulses given by the wave 
equation \cite{SevillaPRE2014}, like signal reverberation \cite{BarrowPhilTransRoySocLond1983}, wake effects \cite{DulaneyPRE2020}, or 
negative probabilities \cite{PorraPRE1997}. In Eq. \eqref{PDFactive-Pulse} $x=\vert\boldsymbol{x}\vert$ and $\Omega_{d}=2\pi^{d/2}/\Gamma(d/2)$ is the 
surface of the $(d-1)$-dimensional unit sphere. 

If we take $d_{\eta}=1$ in \eqref{Pulses}, we recover the propagating pulses given by the solutions of the $d$-dimensional wave equation. As has been 
pointed out in Refs. \cite{BarrowPhilTransRoySocLond1983,SevillaPRE2014,DulaneyPRE2020}, such pulses give rise to wake effects, responsible for the 
anomalous behavior specially in two spatial dimensions where unphysical negative probabilities appear \cite{PorraPRE1997}. Notice additionally that 
the algebraic decay with distance in \eqref{PDFactive-Pulse} differs from the standard decay $x^{-1}$ of the pulse solution of the wave equation in 
three dimensions.

In the long-time regime, $\Gamma t\gg 1$, the dependence on $\eta$ is faded out and the decaying behavior of the memory function \eqref{ActiveKernel} is led by the exponential $e^{-\Gamma(t-t^{\prime})}$, for which normal diffusion is expected. We show that this is the case by use of the approximation $_{1}F_{1}\left(a,b, z\right)\approx \Gamma(a)e^{z}z^{a-b}/\Gamma(b)$ for $z\gg 1$ \cite{Abramowitz1964}, and of $\ell_{p}k\ll 1$; thus the  asymptotic expression of the active sector in the intermediate scattering function (\ref{TotalPDF-Fourier}) behaves as 
\begin{eqnarray}
\widetilde{P}_{\text{a,l-t}}(\boldsymbol{k},t)\simeq \exp[-D_{\rm eff} \boldsymbol{k}^2 t].
\label{asymptoticlt}
\end{eqnarray}
This is exactly the expected expression of the intermediate scattering function corresponding to Brownian motion with an effective diffusion constant $D_{\rm eff}=c^2/\Gamma$ \cite{KurzthalerSciRep2016,MartensEPJE2012}.  

\subsection{\label{subsect:IIb}Mean-squared displacement for the active part of motion}
We focus on the mean-squared displacement associated with the effective transport equation (\ref{GeneralizedEq}) proposed here. The contribution to the MSD due to active motion can be obtained directly  by calculating the integral in (\ref{generalsolMSD}) when $D_{T}=0$:
\begin{eqnarray}
\langle\boldsymbol{x}^2(t)\rangle_{\text{a}}%&=2d\eta!(ct)^{2}\sum_{n=0}^{\infty}\frac{\left(-\Gamma t\right)^{n}}{(n+\eta+1)!(n+2)}\label{MSDseries}\\
 =2d\frac{c^{2}}{\Gamma} t\Bigl[1-\, _{2}F_{2}\bigl(\left\{1,1\right\},\left\{2,d_{\eta}\right\},-\Gamma t\bigr)\Bigr],\label{MSDHyper}
\end{eqnarray}
where $_{2}F_{2}\bigl(\{a,b\},\{c,d\},x\bigr)=\sum_{n=0}\frac{(a)_{n}(b)_{n}}{(c)_{n}(d)_{n}}\frac{x^{n}}{n!}$ is the confluent hypergeometric function \cite{Abramowitz1964} and $\langle F(\boldsymbol{x})(t)\rangle_{\text{a}}$ denotes the expectation value of $F(\boldsymbol{x})$ calculated with the corresponding probability density of active motion $P_{\text{a}}(\boldsymbol{x},t)$. From the series expansion of the confluent hypergeometric function, the ballistic behavior $\langle\boldsymbol{x}^2(t)\rangle_{\text{a}}\simeq 
v_{\text{eff}}^{2}t^{2}(1-\frac{2}{3}\frac{\Gamma t}{\eta}+\ldots)$ is evident in the short-time regime $\Gamma t\ll1$; however, with an effective speed $v_{\text{eff}}$, proportional to $c$, the proportionality factor being the square root of the ratio of the spatial dimensionality to the temporal one, i.e., $v_{\text{eff}}=\sqrt{d/d_\eta}\, 
c$. Thus, the proposed model given by \eqref{GeneralizedEq}, gives rise to the pulse \eqref{Pulses} that propagates with the effective speed $v_{\text{eff}}$. If the spatial dimension $d$ is larger than the temporal one $d_{\eta}$ the effective speed is larger than $c$, and the reversed occurs if $d<d_{\eta}$. As noted above, the case $d_{\eta}=d$ is of particular interest since the parameter $c$ corresponds precisely to the propagation speed.

\begin{figure}
 \includegraphics[width=\columnwidth,clip=true]{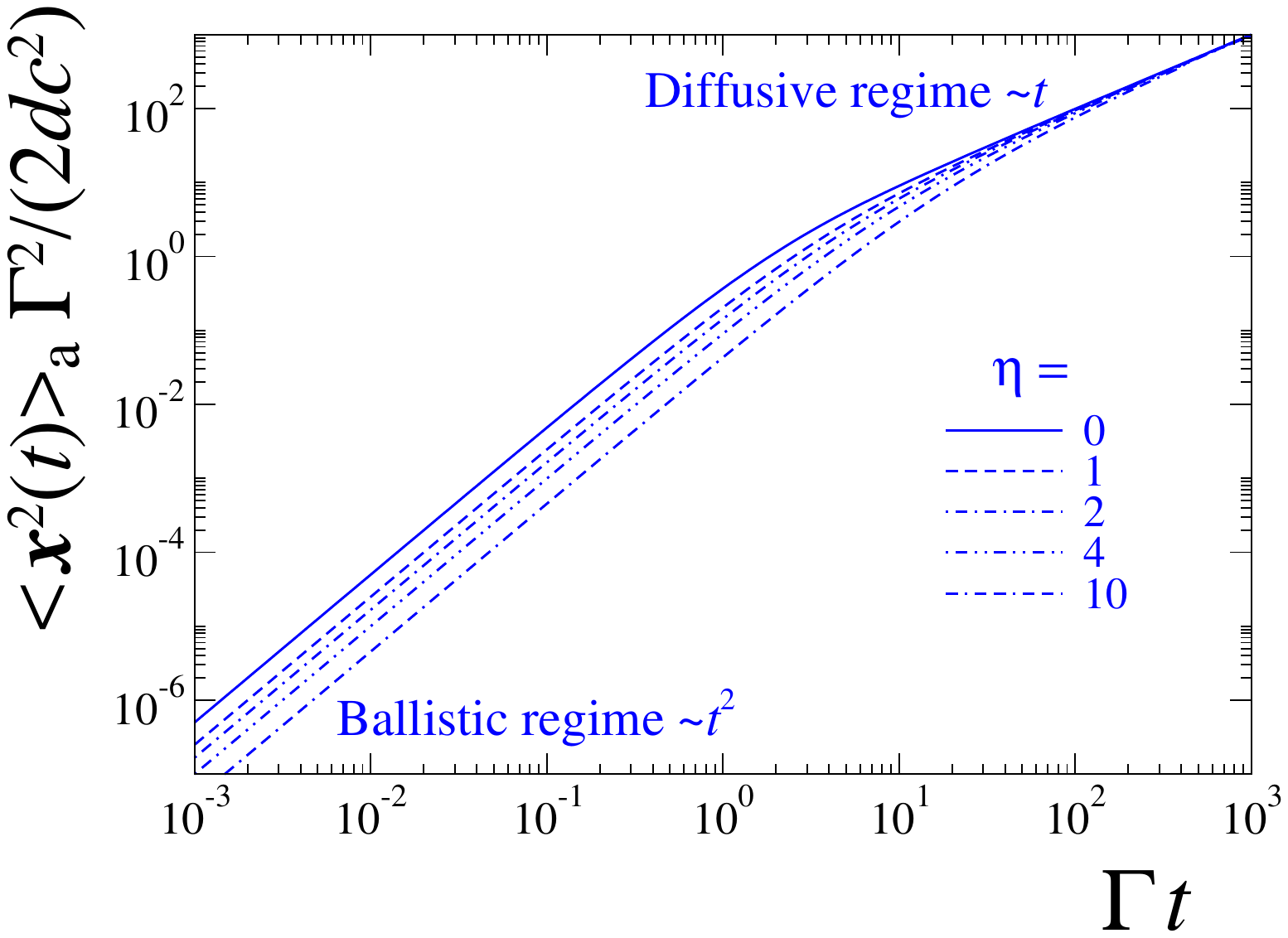} \caption{(Color online) Dimensionless mean-square displacement $\left[\Gamma^{2}/(2dc^{2})\right]\langle\boldsymbol{x}^{2}(t)\rangle_{\text{a}}$ as function of the dimensionless time $\Gamma t$ for different values of $\eta$. The ballistic behavior, $\sim (d/d_{\eta}) c^{2}t^{2}$, is evident in the short-time regime, while normal diffusion $2dD_{\text{eff}}t$ is shown at the long-time regime with $D_{\text{eff}}=c^{2}/\Gamma$.}
\label{Fig-ActiveMSD}
\end{figure}
By use of the asymptotic behavior of the confluent hypergeometric function in (\ref{MSDHyper}), one recovers the diffusive behavior 
\begin{multline}\label{MSDasymptotic}
 \langle\boldsymbol{x}^2(t)\rangle_{\text{a}}=2dD_{\text{eff}}\, t\left\{1+\frac{\eta[\psi(\eta)-\ln\Gamma t]}{\Gamma t}-\frac{\eta(\eta-1)}{(\Gamma t)^{2}}\times\right.\\
\left. \left[1-\frac{\eta-2}{2\Gamma t}+\frac{(\eta-2)(\eta-3)}{3(\Gamma t)^{2}}-\ldots\right]\right\},
\end{multline}
giving rise to normal diffusion $\sim 2dD_{\text{eff}}\, t$  as $\Gamma t \rightarrow\infty$, with the diffusion coefficient $D_{\text{eff}}=c^{2}/\Gamma$, where $\psi(x)$ is the digamma function \cite{Abramowitz1964}. The factor between curly brackets contains the corrections due to the active memory kernel \eqref{ActiveKernel} and thus shows an explicit dependence on $\eta$. Notice that although the model makes $\eta$ conspicuously revealed in the short-time regime, it affects the time dependence of the MSD in the intermediate-time regime as well. That is, while no correction whatsoever to the standard $2dD_{\text{eff}}\, t$ is obtained for $\eta=0$, this linear time dependence of the MSD is corrected by the factor $[1-(\gamma_{\text{e}}+\ln\Gamma t)/\Gamma t]$ for $\eta=1$, with $\gamma_{\text{e}}\simeq0.5772\ldots$ the Euler-Mascheroni constant. Higher corrections appear the higher the value of $\eta$, and these corrections shift the crossover to the diffusive regime to larger times, as is apparent in Fig. \ref{Fig-ActiveMSD}, where the exact time dependence of the MSD is shown 
for different values of $\eta$. We would like to point out in passing, that an equivalent exact expression for the time dependence of the  MSD is given by Eq. \eqref{castro} in Appendix \ref{appC}. This expression is obtained directly from \eqref{Solution} using $\langle\boldsymbol{x}^2(t)\rangle_{\text{a}}=-\nabla^2_{\boldsymbol{k}}\tilde{P}_{\text{a}}(\boldsymbol{k},t)\vert_{\boldsymbol{k}=0}$, where $\nabla^{2}_{\boldsymbol{k}}$ is the Laplacian in the Fourier variables. Expressions \eqref{castro} and \eqref{MSDHyper} establish a mathematical identity that will be proved rigorously elsewhere.

Specific expressions for the MSD obtained from (\ref{MSDHyper}), are given for different values of $\eta$ in Table \ref{tableMSD}. For $\eta=0$, we recover the well-known expression of persistent motion obtained, among many other equations---such as from the Ornstein-Uhlenbeck description of Brownian motion---from the telegrapher's equation. This expression describes the exponentially fast crossover from ballistic motion to normal diffusion, for arbitrary spatial dimension around times of the order of the time-scale $\Gamma^{-1}$.
\begin{table}[h]
\caption{\label{tableMSD}Explicit expressions for the mean-squared displacement from \eqref{MSDHyper} for specific cases of different values of $d_\eta=\eta+1$.}
\begin{ruledtabular}
\begin{tabular}{cc}
 $d_\eta$ & $\langle \boldsymbol{x}^2 (t)\rangle_{\text{a}}$\\\hline
 1  & $\frac{2dc^{2}}{\Gamma^{2}}\left[\Gamma t-\left(1-e^{-\Gamma t}\right)\right]$\\
 2 & $\frac{2dc^{2}}{\Gamma^{2}}\left[\Gamma t-\text{E}_{\text{in}}\left(\Gamma t\right)\right]$,\\
 3 & $\frac{2dc^{2}}{\Gamma^{2}}\left\{\Gamma t+2\left[1-\text{E}_{\text{in}}(\Gamma t)\right]-\frac{2}{\Gamma t}\left(1-e^{-\Gamma t}\right)\right\}$
\end{tabular}
\end{ruledtabular}
$\text{E}_{\text{in}}\left(x\right)=\gamma_{\text{e}}+\ln x+\text{E}_{1}(x)$ where $\text{E}_{1}(x)=\int_{x}^{\infty}dt\, e^{-t}/t$ is the 
exponential integral function (see Ref. \cite{Abramowitz1964}) and $\gamma_{\text{e}}\simeq0.5772\ldots$ is the Euler-Mascheroni constant.
\end{table}

\subsection{The Kurtosis $\kappa(t)$ of $P_{\text{a}}(\boldsymbol{x},t)$}

As is well known, the kurtosis of a probability density function gives information about its ``shape'' and can be used as a measure of ``distance'' from a reference distribution. The kurtosis definition given in Ref. \cite{Mardia74p115} is an invariant measure for multivariate Gaussian distributions and is explicitly given by $ \kappa=\left\langle\boldsymbol{x}^{T}\boldsymbol{\Sigma}^{-1}\boldsymbol{x}\right\rangle$, where $\boldsymbol{x}^{T}$ denotes the transpose of the $d$-dimensional vector $\boldsymbol{x},$ and $\boldsymbol{\Sigma}^{-1}$ the so called covariance matrix defined as the inverse of the average for the dyadic product $\boldsymbol{x}\cdot\boldsymbol{x}^{T}.$ 

For rotationally symmetric distributions, the kurtosis reduces simply to 
\begin{equation}\label{kurtosis}
 \kappa(t)=d^{2}\frac{\left<\boldsymbol{x}^{4}(t)\right>}{\langle\boldsymbol{x}^{2}(t)\rangle^{2}}.
\end{equation}
It can be shown that the kurtosis for a rotationally symmetric Gaussian distribution with mean $\boldsymbol{\mu}$ and variance $\sigma^{2}$ has the invariant kurtosis value 
$d(d+2)$. Thus, in the long-time regime it is expected that the kurtosis of the distributions given by \eqref{TotalPDF} and \eqref{Solution} acquires this value, denoted with $\kappa_{\infty}$.

\begin{figure}[t]
\includegraphics[width=\columnwidth]{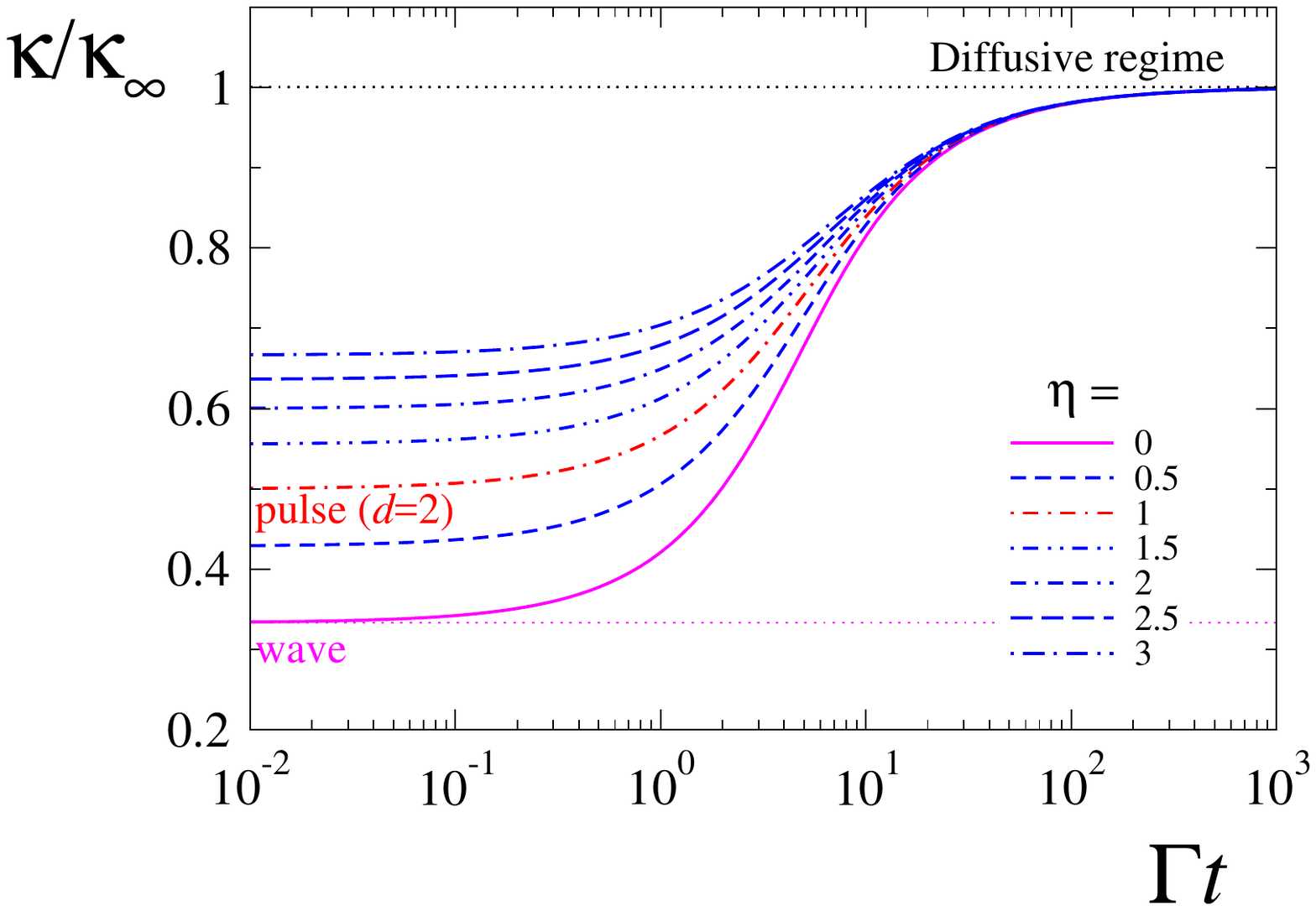}
 \caption{(Color online) Kurtosis $\kappa/\kappa_{\infty}$ as function of the dimensionless time $\Gamma t$ for different values of the parameter $\eta$, where $\kappa_{\infty}=d(d+2)$ gives the value of the kurtosis for a Gaussian distribution in $d$ spatial dimensions. The dash-dotted line marks the case $\eta=1$ for which $d_{\eta}=2$. The value 0.5 in the short-time limit corresponds to the kurtosis of the pulse given by \eqref{PDFactive-Pulse} with $d=2$. }
 \label{Fig-kurtosis}
\end{figure}
  
An exact expression for the time dependence of $\left\langle\boldsymbol{x}^{4}(t)\right\rangle_{\text{a}}$, suitable for analyzing its leading dependence in the short-time regime, is obtained by calculating the integral in (\ref{generalsol4moment}) when $D_{T}=0$. Although it is possible to express it in terms of hypergeometric functions, the following series expansion in powers of $\Gamma t$, 
\begin{multline}
 \langle\boldsymbol{x}^{4}(t)\rangle_{\text{a}}=8d(d+2)\eta!(ct)^{4}\sum_{n=0}^{\infty}\frac{(-\Gamma t)^{n}}{(n+\eta+3)!}\\
\times\frac{n+1+\eta\bigl[\gamma_{\text{e}}-1+\psi(n+3)\bigr]}{n+4},
\end{multline}
is more convenient for the sake of analysis in the short-time regime. In this regime the probability density function depends strongly on $\eta$, as can be seen from the last expression, since we have for $\Gamma t\ll1$ that $\langle\boldsymbol{x}^{4}(t)\rangle_{\text{a}}\simeq d(d+2)/((\eta+3)(\eta+1)) (ct)^{4}$; and after dividing this by the mean-squared displacement in the same time regime, we have that the kurtosis has the $\eta$-dependent value $\kappa_{\text{a-str}}\simeq \kappa_{\infty}(\eta+1)/(\eta+3)$, where $\kappa_{\infty}=d(d+2)$ and $\kappa_{\text{a-str}}$ denotes the kurtosis $\kappa_{a}(t)$ of the probability density $P_{\text{a}}(\boldsymbol{x},t)$ in the short-time regime. In Fig. \ref{Fig-kurtosis} the ratio 
$\kappa_{\text{a}}(t)/\kappa_{\infty}$, which is independent of the system spatial dimensionality, is shown as a function of the dimensionless time $\Gamma t$ for different values of $\eta$. 

Notice that the value $\eta=0$ corresponds to the case described by the telegrapher's equation \cite{SevillaPRE2014} and is characterized by $\kappa_{\text{a-str}}/\kappa_{\infty}=1/3$. As noted in Refs. \cite{SevillaPRE2014,SevillaPRE2016}, for $d=2$, and 3, $\kappa_{\text{a-str}}$ describes a wavelike front propagation for which wake effects are present, leading to the values $\kappa_{\text{a-str}}=8/3$ and $\kappa_{\text{a-str}}=5$ for two and three dimensions, respectively. Likewise, the connection previously devised between 
$\eta$ and the system spatial dimensionality, $\eta=d-1$, which characterizes the short-time propagating pulse \eqref{PDFactive-Pulse}, gives 
$\kappa_{\text{a-str}}/\kappa_{\infty}=d/(d+2)$, for which the value $1/2$ is obtained for $d=2$ (see dash-dotted red line in Fig. \ref{Fig-kurtosis}). Thus, different front propagations are characterized by the ratio $\kappa_{\text{a-str}}/\kappa_{\infty}.$ 

In the long-time regime, the fourth moment of $P_{\text{a}}(\boldsymbol{x},t)$, can be obtained alternatively from \eqref{Solution} by use of the formula $\langle\boldsymbol{x}^4(t)\rangle_{\text{a}}=\nabla^4_{\boldsymbol{k}}\tilde{P}_{\text{a}}(\boldsymbol{k},t)\vert_{\boldsymbol{k}=0}$ (see Appendix \ref{appC}), where $\nabla^{4}_{\boldsymbol{k}}$ is the bi-Laplacian in the Fourier variables. After a long but straightforward calculation, one gets expression (\ref{eq4mom}) for the fourth moment, and from this one, the ratio $\kappa_{\text{a}}/\kappa_{\infty}=1$, which characterizes the diffusive regime of the probability density distribution, is obtained in the long-time regime (see Fig. \ref{Fig-kurtosis} upper dotted line).

\section{\label{Sect-III}Comparison: Generalized Active Brownian Motion \emph{vs} Active Brownian motion, and Run-and-tumble Motion}

In this section we focus our analysis on a comparison of our model of active motion, which arises from Eqs. \eqref{continuity-active} and \eqref{ActiveCurrent}, with the well-known models of active Brownian and run-and-tumble motion. Although the formalism introduced here to consider thermal fluctuations is well applicable to any kind of active motion, we center our attention on discerning the differences between these just in their active nature.

In particular it is of interest to contrast the results obtained from the model we have named generalized active Brownian motion (GABM) described in Sec. \ref{Sect-II},  with the theoretical predictions made for active agents that use the reorientation dynamics of the swimming direction of ABM and RTM. Both models can be cast into a transport equation for the probability density function, $P(\boldsymbol{x}, \boldsymbol{\hat{v}}, t)$, of finding a particle at the point $\boldsymbol{x}$ moving along the swimming direction $\hat{\boldsymbol{v}}$ at the instant $t$. When the translational motion of the active particle occurs in the $d$-dimensional Euclidean space, the dynamics of the swimming vector velocity occurs on the surface of the unit $(d-1)$-dimensional sphere $S^{d-1}$. 

The probability density function for active Brownian motion, $P_{\text{ABM}}(\boldsymbol{x},\hat{\boldsymbol{v}},t)$, satisfies the Fokker-Planck-like
equation
\begin{multline}
\frac{\partial}{\partial t}P_{\text{ABM}}(\boldsymbol{x},\boldsymbol{\hat{ v}}, t)+v_{0}\boldsymbol{\hat{v}}\cdot\nabla P_{\text{ABM}}(\boldsymbol{x},\boldsymbol{\hat{ v}}, t)=\\
D_{\text{R}}\nabla^2_{\hat{\boldsymbol{v}}}P_{\text{ABM}}(\boldsymbol{x},\boldsymbol{\hat{v}}, t),
\label{ABM}
\end{multline}
where $D_{\text{R}}$ is the coefficient of rotational diffusion whose inverse, $D_{\text{R}}^{-1}$, defines the persistence time, $v_{0}$ is the particle swimming speed, $\nabla^2_{\hat{\boldsymbol{v}}}$ is the corresponding Laplace-Beltrami operator associated to the unit sphere $S^{d-1}$, and $\nabla$ is the gradient operator in $d$ spatial dimensions. Analogously, the probability density function for RTM model, $P_{\text{RTM}}(\boldsymbol{x},\boldsymbol{\hat{ v}}, t)$, satisfies the transport equation
\begin{multline}
\frac{\partial}{\partial t}P_{\text{RTM}}(\boldsymbol{x},\boldsymbol{\hat{ v}}, t)+v_{0}\boldsymbol{\hat{v}}\cdot\nabla P_{\text{RTM}}(\boldsymbol{x},\boldsymbol{\hat{ v}}, t)=\\
-\lambda P_{\text{RTM}}(\boldsymbol{x},\boldsymbol{\hat{ v}}, t) + \lambda\int \frac{d\boldsymbol{\hat{v}}^{\prime}}{\Omega_{d}} P_{\text{RTM}}(\boldsymbol{x},\boldsymbol{\hat{v}}^{\prime}, t),
\label{RT}
\end{multline}
where the parameter $\lambda$ is the tumbling rate of the self-propelling particle and $\int d\boldsymbol{\hat{v}}^{\prime}$ denotes the integration over the $d$-dimensional unit sphere $S^{d-1}$, $\Omega_{d}$ being its area.  A comparison between ABM and RTM has been carried out theoretically to study the many-body effects of active particles that exhibit motility-induced phase separation in two-dimensions \cite{CatesEPL2013}. On the other hand, a comparison between both models against data acquired from trajectories of Janus particles has been carried out in two dimensions too \cite{KurzthalerPRL2018}, where the authors conclude that the model of active Brownian motion describes better the data analyzed. 

Exact analytical expressions for the total intermediate scattering function are known for active Brownian particles freely diffusing in two \cite{KurzthalerPRL2018} and three dimensions \cite{KurzthalerSciRep2016}, while the corresponding active intermediate scattering functions in one, two, and three dimensions for RTM are given in Ref. \cite{MartensEPJE2012}. 

In addition to a direct comparison between the ISF of ABM and RTM with the one given in this paper, [see Eq. \eqref{TotalPDF-Fourier}], we make a comparative analysis among the mean-squared displacements and the kurtosis of the active part of the particle's motion. This allows for a more physical analysis of the differences among these patterns of active 
motion. 

For ABM and RTM the mean-squared displacement and the kurtosis are calculated from an analysis based on terms of the hydrodynamic-like tensor fields defined in the Appendix \ref{AppendixActiveMotion}, namely, $\rho(\boldsymbol{x},t)$, the scalar probability density \eqref{Def-rho}; $\mathbb{P}^{i}(\boldsymbol{x},t)$, the $i$th component of the polarization vector field \eqref{Def_Polarization}; $\mathbb{Q}^{ij}(\boldsymbol{x},t)$, the elements of the nematic tensor field \eqref{Def-Nematic}; and the elements of the third rank tensor field $\mathbb{R}^{ijk}(\boldsymbol{x},t)$ \eqref{Def_Rij}. 

The procedure presented in the Appendix (\ref{AppendixActiveMotion}) allows us to derive equations for each of the hydrodynamic-like tensors just mentioned above. The conservation of probability leads to the continuity equation
\begin{eqnarray}
\frac{\partial}{\partial t}\rho(\boldsymbol{x}, t)+\partial_{i}J^{i}(\boldsymbol{x},t)=0,
\label{eqRho}
\end{eqnarray}
where the $i$th component of the probability current, $J^{i}(\boldsymbol{x},t)$, is given by $v_{0}\mathbb{P}^{i}(\boldsymbol{x},t)$ and $\partial_{i}$ denotes the partial derivative with respect to the $i$th spatial coordinate. 
Likewise, the evolution equation of the $i$th component of the polarization vector $\mathbb{P}^{i}(\boldsymbol{x},t)$ is 
\begin{equation}
\frac{\partial}{\partial t} \mathbb{P}^{j} =-\xi\, \mathbb{P}^{j}-\frac{v_{0}}{d}\partial_{j}\rho-v_{0}\partial_{k}\mathbb{Q}^{jk},
\label{eqPol}
\end{equation}
where the parameter $\xi$ acquires the value $(d-1)D_{\text{R}}$ for the ABM model, and the value $\lambda$ for the RTM one, respectively. We note that by combining Eqs. (\ref{eqRho}) and (\ref{eqPol}),  one gets the following inhomogeneous telegrapher's equation: \begin{eqnarray}
\frac{\partial^2}{\partial t^2}\rho+\xi\frac{\partial}{\partial t} \rho=\frac{v^2_{0}}{d}\nabla^2\rho+v_{0}^2\partial_{i}\partial_{j}\mathbb{Q}^{ij}.
\label{MTE1}
\end{eqnarray}
On the other hand, the evolution equation for the elements of the nematic-like tensor $\mathbb{Q}^{ij}$ IS given by   
\begin{eqnarray}
\frac{\partial }{\partial t}\mathbb{Q}^{ij}&=&-\overline{\xi} \mathbb{Q}^{ij}-\frac{v_{0}}{d+2}\mathbb{T}^{ij}-v_{0}\partial_{k}\mathbb{R}^{ijk},
\label{EqQ}
\end{eqnarray}
with $\mathbb{T}^{ij}=-2\frac{\delta^{ij}}{d}\partial_{k}\mathbb{P}^{k}+\partial^{i}\mathbb{P}^{j}+\partial^{j}\mathbb{P}^{i}$ a second 2-rank traceless tensor field, and the parameter $\overline{\xi}$ acquires the value $2dD_{R}$ for ABM and the value $\lambda$ for the RTM, respectively.  

Notice that the damping rate in Eq. \eqref{eqPol} (first term in the right-hand side) is given by $\xi$, while it is given by $\overline{\xi}$ in Eq. \eqref{EqQ}. Thus, $\mathbb{P}^{i}$ and $\mathbb{Q}^{ij}$ damp out at the same rate for the case of RTM, since $\xi=\overline{\xi}=\lambda$. In contrast, the nematic tensor $\mathbb{Q}^{ij}$ damps out faster than the polarization vector $\mathbb{P}^{i}$ for ABM, since $\xi>\overline{\xi}$ for $d\ge2$. This is expected to occur also for tensors of higher rank, namely, that each tensor of rank $n+1$ decays faster than the one of rank $n$. This observation demonstrates the difference between both models; in particular this is analyzed by a comparison of the time dependence of the kurtosis among the models.

\subsection{The mean-squared displacement and the kurtosis}
\begin{figure}
\includegraphics[width=\columnwidth, trim= 30 50 0 10]{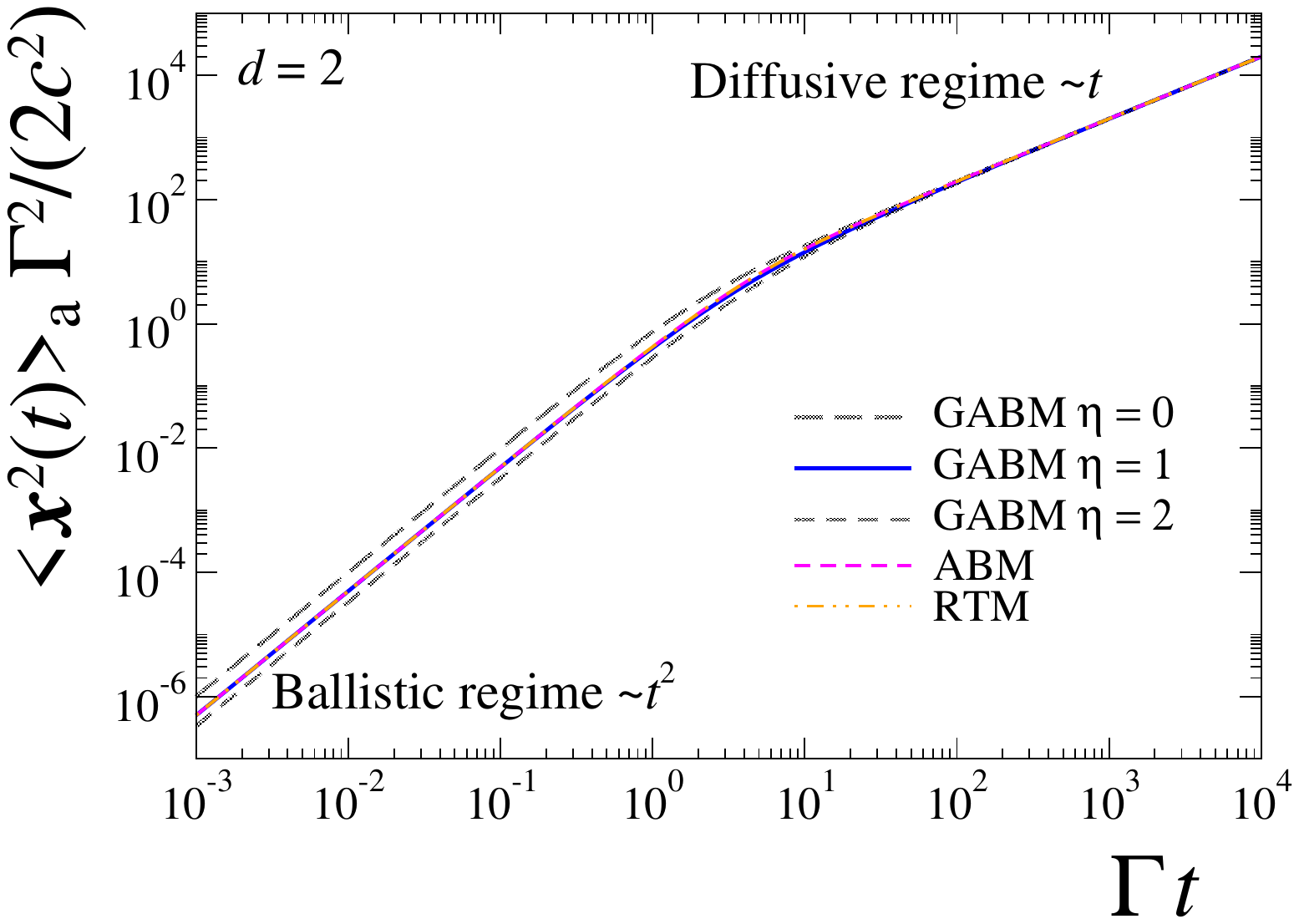}
 \caption{(Color online) Comparison of the time dependence of mean-squared displacement given by the GABM \eqref{MSDHyper} ($\eta=0$, 1, and 2) and the mean-squared displacement corresponding to ABM and RTM \eqref{MSD-Paradigmatic}, for the two-dimensional case ($d=2$).} \label{MSD1o}
\end{figure}
The exact time dependence of the mean-squared displacement for ABM and RTM can be obtained straightforwardly from the inhomogeneous telegrapher's equation [as is shown in Appendix \ref{AppendixActiveMotion} by use of Eq. \eqref{MTE1}] and is given by  
 \begin{equation}\label{MSD-Paradigmatic}
\langle\boldsymbol{x}^2(t)\rangle_\text{a}=\frac{2v_{0}^2}{\xi^2}\left[\xi t-(1-e^{-\xi t})\right].
\end{equation}
As has been mentioned in Sec. \ref{subsect:IIb}, this expression arises from a variety of models of persistent motion, exhibiting the ballistic time dependence $\langle\boldsymbol{x}^2(t)\rangle\simeq v^2_{0}t^2$, in the short-time regime $\xi t\ll 1$, and the linear time dependence $\langle\boldsymbol{x}^2(t)\rangle\simeq 2d D_{\text{eff}}\, t$, in the regime of long times, $\xi t\gg 1$, with $D_{\text{eff}}=v_{0}^{2}/(d\,\xi)$. For the GABM model presented in this paper,  Eq. \eqref{MSD-Paradigmatic} arises naturally only for the case $\eta=0$ (see Table \ref{tableMSD}). Notice, however, that there is no quantitative agreement between the MSD for $\eta=0$ and the expression \eqref{MSD-Paradigmatic} \emph{for constant values of the set of parameters}: \{$c$, $\Gamma$, $v_{0}$, $D_{\text{R}}$ and $\lambda$\} (see the curves for $\eta=0$, long-dashed gray line, ABM dashed-magenta line and RTM dashed-doble-dotted orange line in Fig. \ref{MSD1o} for the two-dimensional case). Nevertheless, a good agreement among the time dependence of the three models (see the solid-blue line in Fig. \ref{MSD1o} for two spatial dimensions and the lines for ABM and RTM) is obtained when the spatial dimensionality, $d$, plays the same role as the temporal one, i.e., when $d_{\eta}=d$ or equivalently $\eta=d-1$. In addition we must chose $\Gamma=D_{R}d(d-1)$ for a proper comparison with ABM and $\Gamma=\lambda d$ to compare with RTM. In both cases we set $c=v_{0}$. That such is the case can be easily proof by requiring that the moments at both the short- and long-time regimes of our model, coincide with the corresponding ones of the ABM and RTM.

\begin{figure}
\includegraphics[width=\columnwidth, trim= 30 50 0 10]{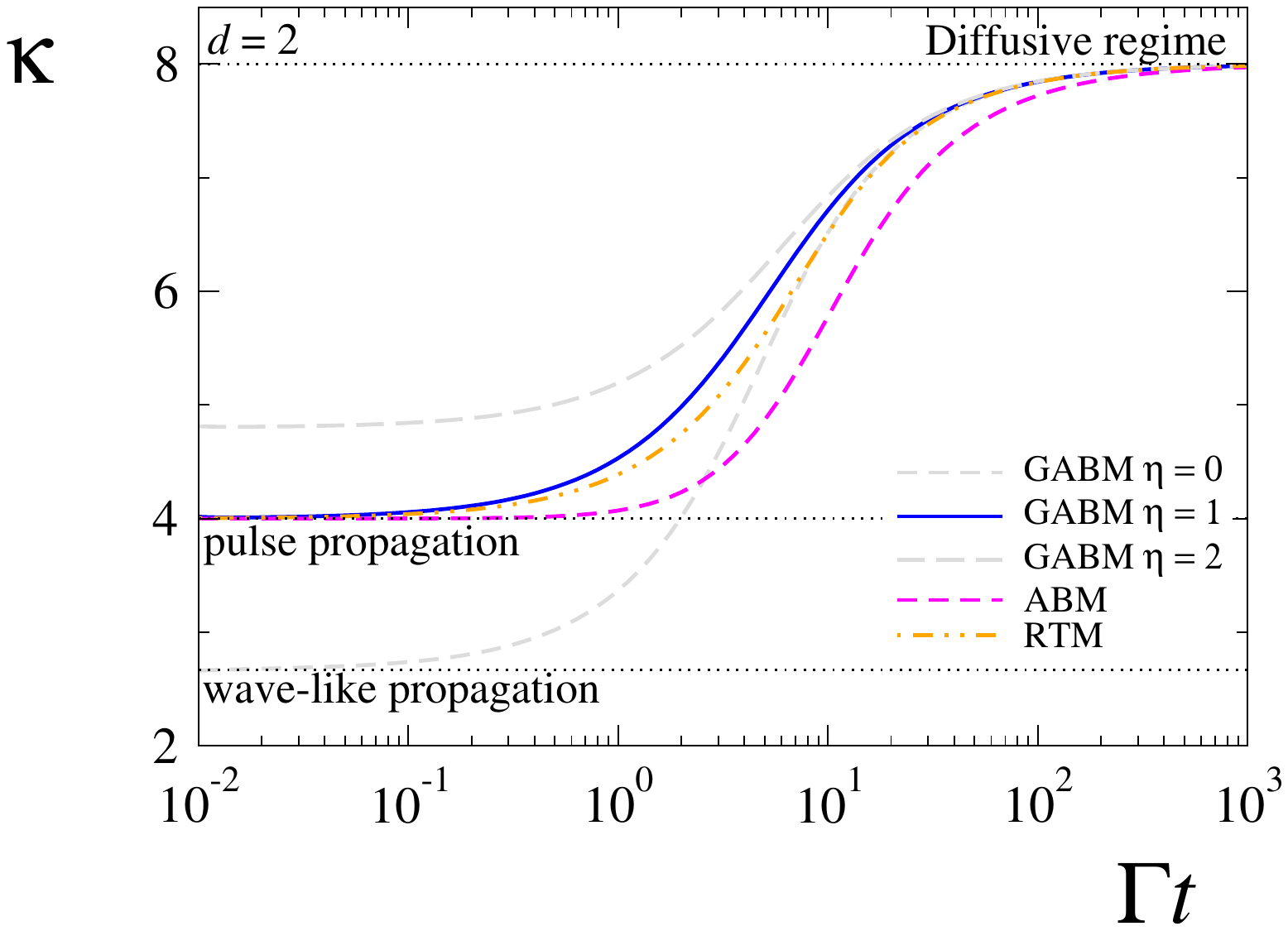}
 \caption{(Color online) Comparison of the time dependence of the probability distribution kurtosis given by the GABM for $\eta=0$ (dashed-gray line), 1 (solid-blue line), and 2 (short-dsashed gray line),  and the mean-squared displacement corresponding to ABM and RTM \eqref{MSD-Paradigmatic}, for the two-dimensional case ($d=2$).} \label{KurtosisComparisson}
\end{figure}
We want to point out that even when the agreement in the mean-squared displacement is good when we choose $\eta=d-1$, the position distribution reveals differences among the three patterns of active motion considered. This is markedly exhibited in the intermediate-time regime of the kurtosis shown in Fig.~\ref{KurtosisComparisson} for $d=2$, where for the timescales chosen, the GABM grows more rapidly than ABM and RTM. In the same figure it can be noticed that in the short-time  regime, the case for which $\eta=1$, properly describes the distorsionless pulse propagation (characterized by $\kappa=4$ in two spatial dimensions) exhibited by ABM and RTM in contrast to the case $\eta=0$, which gives rise to wake effects of 
wavelike propagation characterized by $\kappa=8/3\simeq 2.667$ in two spatial dimensions \cite{SevillaPRE2014}. In the long-time regime the kurtosis of the three distributions reaches the value 8 that univocally characterizes the Gaussian distribution of normal diffusion.

\subsection{The Intermediate Scattering Function: Two-dimensional case}
\begin{figure*}[t]
\includegraphics[trim=16 120 25 45,clip=true,width=\textwidth]{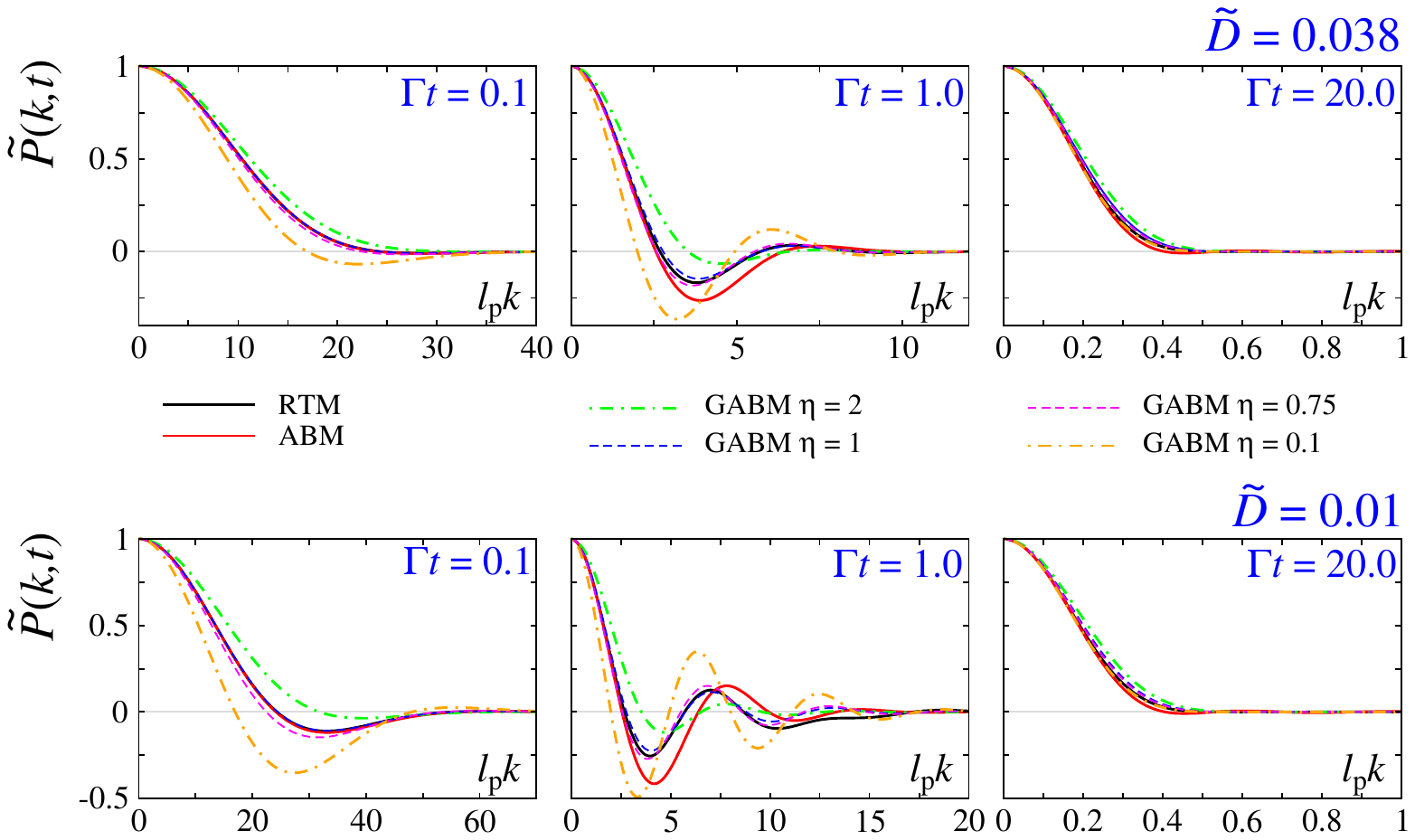}
\caption{(Color online) Intermediate scattering function $\widetilde{P}(\boldsymbol{k},t)$ given by \eqref{Solution} \emph{vs} the corresponding ISF obtained numerically as explained in the Appendix \ref{AppendixABM-RTM}. The first row for $\widetilde{D}=D_\text{T}/\Gamma\ell^{2}_\text{p}=0.038$ is explained in Fig. \ref{PDFk_t}, and the second corresponds to the value $\widetilde{D}=0.01$ for which the effects of active motion are more conspicuous, particularly at times of the order of $\Gamma^{-1}$, i.e. $\Gamma t\sim 1$; see central panels.}
\label{ISF-Comparison}
\end{figure*}

Finally for this section, we compare the ISF given by our model of active motion for $\eta>0$, Eq. \eqref{TotalPDF-Fourier}, with the corresponding complete ISF of ABM and RTM in the case of two dimensions. These last ones are computed by taking the product of $\exp\{-D_\text{T}k^{2}t\}$ and the Fourier transform of $P_\text{ABM}(\boldsymbol{x},t)$ and $P_\text{RTM}(\boldsymbol{x},t)$ obtained by integrating over the orientational degrees of freedom of $P_\text{ABM}(\boldsymbol{x},\hat{\boldsymbol{v}},t)$ and $P_\text{RTM}(\boldsymbol{x},\hat{\boldsymbol{v}},t)$, respectively. Instead of using the expansions in eigefunctions to compute the ISFs, as is done in Ref. \cite{KurzthalerPRL2018}, we use instead the exact representation of these in the form of continuous fractions which are amenable for numerical evaluation (see Appendix \ref{AppendixABM-RTM}).

Our results are shown in Fig. \ref{ISF-Comparison} for two different values of the inverse P\'eclet number, $\widetilde{D}=0.038$ (such as the one used in Fig. \ref{PDFk_t}), and $\widetilde{D}=0.01$ for which the effects of persistence are more conspicuous. The ISF obtained from our model as function of the dimensionless wave vector $\ell_{p}k$ is shown for $\eta=2$ (dash-dotted green line), 1 (dashed blue line), 0.75 (dashed magenta line) and 0.1 (dash-dotted magenta line) is compared with the corresponding ones of RTM (solid black line) and of ABM (solid red line).

\section{Concluding remarks and perspectives}\label{Sec. IV}

In this paper we proposed and studied a model for active motion that incorporates the persistence effects of self-propulsion through the memory function given in Eq. \eqref{ActiveKernel}. The specific choice of such function results in the Smoluchowski-like equation \eqref{GeneralizedEq} for the probability density function $P_{\text{a}}(\boldsymbol{x},t)$ of finding a particle at the position $\boldsymbol{x}$ at the time $t$ in a $d$-dimensional space. The main feature of the memory kernel \eqref{ActiveKernel} is that it gives a smaller weight to events in the past than the corresponding exponential memory of the telegrapher process \cite{ForsterBook, DunkelPhysRep2009}. This leads to Eq. \eqref{GeneralizedEq}, which corresponds to an original generalization of the well-known telegrapher's equation. Such a generalization takes into consideration, in an exact manner, the contribution of active motion characterized by the parameter $\eta$, which carries implicitly, the mechanism of self-propulsion.
Our analysis takes into consideration the fluctuations of a thermal bath as well. The joint effects of thermal and active fluctuations are encompassed in the total probability density function $P(\boldsymbol{x},t)$, given by Eq. \eqref{TotalPDF}, whose Fourier transform gives us the intermediate scattering function [see Eq. \eqref{TotalPDF-Fourier}]. Thus, our analysis avoids the sometimes cumbersome hierarchy analysis made for active Brownian and run-and-tumble motion and allows a qualitative comparison with the experimental data for Janus particles \cite{KurzthalerPRL2018}.

We also have presented an analysis of the time dependence of the mean-squared displacement and of the kurtosis of the particle position distribution, and we studied the deep consequences of the memory parameter $\eta$. To be explicit, we uncovered a large class of active-motion patterns that propagate distinctly as the parameter $\eta$ is varied. The connection between $\eta$ and the spatial dimensionality $d$ is uncovered by choosing $\eta=d-1$; for this, the ``time dimension'' plays the same role as the spatial dimension in the context of the generalization of the wave equation of Bietenholz and Gambiagi. In our model, this choice manifestly captures the dynamics of active Brownian motion and run-and-tumble motion, if the relations $\Gamma=D_{R}d(d-1)$ and $\Gamma=\lambda d$ are satisfied.

The approach presented in this paper can be extended in various directions. For instance, it is of interest to consider the analysis  of the combined effects of general patterns of active motion under the confinement effects of an external field. This case can be treated directly by considering an extra term in the current probability due to the external force. In addition, the effects of spatial heterogeneity on the dynamics of the active particle, such as when the particle motion occurs on a curved surface, can be immediately generalized for the GABM. Moreover, as we mentioned above, the memory function can be adjusted in order to capture much more different active behaviors. For example, we can modify the time dependence of the memory function in order to damp it out slower than the function proposed here. Also, the memory function can be modified to take into account a space dependence to describe additional effects of spatial inhomogeneities. Furthermore, we can use the space dependence of the memory function to interpret it as a generalized diffusion process in order to treat the problem of an interacting active particle system within the context of the hydrodynamic fluctuation theory \cite{ForsterBook, HessKlein}.  

\begin{acknowledgments}
This work was supported by UNAM PAPIIT-IN110120 and UNAM-PAPIIT-IN114717. P.C.V. acknowledges financial support by CONACyT Grant No. 237425. 
\end{acknowledgments}
\appendix 

\section{\label{sect:Proof}Consistency between constitutive relations  \eqref{TotalCurrent} and \eqref{TotalPDF} with the active continuity equation (\ref{continuity-active}).}

Here we give a proof of the statement given in Sec. \ref{Sect-II}, that Eqs. \eqref{TotalCurrent} and \eqref{TotalPDF}  provide a ``constitutive relation'' for the continuity equation \eqref{continuity} and are consistent with the active continuity equation (\ref{continuity-active}). Indeed, by direct substitution of Eqs. \eqref{TotalCurrent} and \eqref{TotalPDF} in \eqref{continuity} we have, after some rearrangements, that
\begin{multline}
 \int d\boldsymbol{x}^{\prime}\left[\frac{\partial}{\partial t}G(\boldsymbol{x}-\boldsymbol{x}^{\prime}, t)\right]P_\text{a}(\boldsymbol{x}^\prime, 
t)\\
+\int d\boldsymbol{x}^{\prime}G(\boldsymbol{x}-\boldsymbol{x}^{\prime}, 
t)\frac{\partial}{\partial t}P_\text{a}(\boldsymbol{x}^\prime,t)\\
 -\int d\boldsymbol{x}^{\prime}D_{T}\left[\nabla^2 G(\boldsymbol{x}-\boldsymbol{x}^{\prime}, 
t)\right]P_{a}\left(\boldsymbol{x}^{\prime}, t\right)\\
+\int d\boldsymbol{x}^{\prime}\left[\nabla G(\boldsymbol{x}-\boldsymbol{x}^{\prime}, t)\right]\cdot\boldsymbol{J}_\text{a}(\boldsymbol{x}^\prime, 
t)=0
\end{multline}
The first and third terms of the last equation cancel each other by use of the diffusion equation for the Brownian propagator, 
\begin{eqnarray}
 \frac{\partial}{\partial t}G(\boldsymbol{x}-\boldsymbol{x}^{\prime}, t)=D_{T}\nabla^2 G(\boldsymbol{x}-\boldsymbol{x}^{\prime}, t).\label{eqB}
\end{eqnarray}
We get
\begin{multline}
\int d\boldsymbol{x}^{\prime}\biggl[G(\boldsymbol{x}-\boldsymbol{x}^{\prime}, 
t)\frac{\partial}{\partial t}P_\text{a}(\boldsymbol{x}^\prime,t)
 +\\
 [\nabla G(\boldsymbol{x}-\boldsymbol{x}^{\prime}, t)]\cdot\boldsymbol{J}_\text{a}(\boldsymbol{x}^\prime, 
t)\biggr]=0,
\end{multline}
Now, we use the identity $\nabla G(\boldsymbol{x}-\boldsymbol{x}^{\prime}, t)=-\nabla^{\prime} G(\boldsymbol{x}-\boldsymbol{x}^{\prime}, t)$ where $\nabla^{\prime}$ is the gradient operator in the ${\boldsymbol{x}^{\prime}}$ variables, and perform an integration by parts to have
\begin{equation}
       \int d\boldsymbol{x}^{\prime}G(\boldsymbol{x}-\boldsymbol{x}^{\prime},t)\biggl[\frac{\partial}{\partial 
t}P_\text{a}(\boldsymbol{x}^\prime,t)+\nabla^{\prime}\cdot\boldsymbol{J}_\text{a}(\boldsymbol{x}^{\prime},t)\biggr]=0,
      \end{equation}
from which we obtain Eq. \eqref{continuity-active}, which is completed by Eq. \eqref{ActiveCurrent}.

\section{\label{appDiff}Solution of the differential equation \eqref{TransformedGeneralizedEqFourier}}

In order to solve the Eq. (\ref{TransformedGeneralizedEqFourier}), we make the substitution $\widetilde{P}_\text{a}(\boldsymbol{k}, t)=e^{-x(p)T}\varphi({\bf p}, T)$ where $x(p)=\frac{1}{2}(1+\sqrt{1-4p^2})$, where $T=\Gamma t$, $p=\vert\boldsymbol{p}\vert$, and $\boldsymbol{p}=\boldsymbol{k}\ell_{p}$ are dimensionless quantities introduced here for the sake of clarity. In addition, we make the change of variable $\tau=T\bigl(2x(p)-1\bigr)$; thus it is straightforward to show that the function $\varphi(\boldsymbol{p},\tau)$ satisfies the Kummer differential equation 
 \begin{eqnarray}
\tau\frac{d^2\varphi}{d\tau^2}+\left(\eta-\tau\right)\frac{d\varphi}{d\tau}-\alpha\varphi=0\nonumber, 
\end{eqnarray}
where $\alpha=\eta x(p)/(2x(p)-1)$. It is well known  that this differential equation has the two linearly independent solutions $\Phi(\alpha,\eta;\tau)$ and $\tau^{1-\eta}\Phi(\alpha-\eta+1,2-\eta; \tau)$, where $\Phi(\alpha,\eta; z)={_{1}F_{1}}(\alpha,\eta; z)$ is the confluent hypergeometric function \cite{Abramowitz1964}.  The first solution, $\Phi(\alpha,\eta;\tau)$, is the only possible one that satisfies the initial conditions $\widetilde{P}_\text{a}(\boldsymbol{k}, 0)=1$ and $d\widetilde{P}_\text{a}(\boldsymbol{k }, 0)/dT=0$, unless the parameter $\eta=0$. In such a case, both linear independent solutions must be considered; thus the solution is of the form
\begin{eqnarray}
\widetilde{P}_\text{a}(\boldsymbol{k}, 0)&=&e^{-x(p)T}\lim_{\eta\to 0}\left[A_{p}\Phi(\alpha,\eta;\tau)\right.\nonumber\\&+&\left.B_{p}\tau^{1-\eta}\Phi(\alpha-\eta+1,2-\eta; \tau)\right].
\end{eqnarray}
Now, using that $\lim_{\eta\to 0}\Phi(\alpha,\eta;\tau)=\Phi(0,0,\tau)=1$,  $\lim_{\eta\to 0}\tau^{1-\eta}\Phi(\alpha-\eta+1,2-\eta; \tau)]=\tau\Phi(1,2; \tau)]=e^{\tau}-1$, and after imposing the initial value conditions just mentioned above, it is not difficult to show that $A_{p}=1$ and $B_{p}=x(p)/(2x(p)-1)$.  After a straightforward calculation the solution becomes
\begin{multline}
\widetilde{P}_\text{a}(\boldsymbol{k}, t)
=e^{-\frac{1}{2}\Gamma t}\left[\cosh\left(\frac{\Gamma t}{2}\sqrt{1-4\ell_{p}^2 k^2}\right)\right.\nonumber\\+\left.\frac{\sinh\left(\frac{\Gamma t}{2}\sqrt{1-4\ell_{p}^2 p^2}\right)}{\sqrt{1-4\ell_{p}^2 p^2}}\right],
\end{multline}
which is the well-known solution for the telegrapher equation (\ref{TelegrapherEq}) \cite{SevillaPRE2015}.

\section{Alternative expressions for MSD and Kurtosis}\label{appC}

In this section,  we give the expressions for the mean-squared displacement and fourth moment directly from the probability distribution function (\ref{Solution}). In the case of MSD one gets
\begin{eqnarray}
\langle \boldsymbol{x}^2(t)\rangle_{\text{a}}=2d\left[\frac{c^{2}}{\Gamma}t-\eta e^{-\Gamma t}\sum_{k=1}^{\infty}\frac{(\Gamma t)^{k}}{k!}\left[\psi(\eta+k)-\psi(\eta)\right]\right],\nonumber\\
\label{castro}
\end{eqnarray}
where $\psi(x)=d\log\Gamma\left(x\right)/dx$ is the digamma function.  The appearance of the exponential factor in front of the sum makes expression (\ref{castro}) useful to determine the asymptotic regime, that is, $\langle\boldsymbol{x}^2\rangle_{\text{a}}\simeq 2d\, D_{\text{eff}}t$, which is the standard expression in the diffusive regime, where the effective diffusion constant is given by $D_{\rm eff}=c^{2}/\Gamma$.   In addition, we obtain the following expression for the time dependence of the fourth moment:
\begin{eqnarray}
\langle\boldsymbol{x}^{4}\left(t\right)\rangle_{\text{a}}&=&\left.\frac{d(d+2)c^4}{\Gamma^4}\right.\left\{-8\Gamma t+4\Gamma^2t^2\nonumber\right.\\&+&\left.8(\Gamma t+3)\eta ~e^{-\Gamma t}
\sum_{k=1}^{\infty}\frac{(\Gamma t)^{k}}{k!}\left(\psi(\eta+k)-\psi(\eta)\right)\right.\nonumber\\
&-&\left.16\eta~ \Gamma t e^{-\Gamma t}\sum_{k=1}^{\infty}\frac{(\Gamma t)^{k-1}}{(k-1)!}\left(\psi(\eta+k)-\psi(\eta)\right)\nonumber\right.\\
&+&\left.4\eta^2e^{-\Gamma t}\sum_{k=1}^{\infty}\frac{(\Gamma t)^{k}}{k!}\left(\psi(\eta+k)-\psi(\eta))^2\right]\right.\nonumber\\
&+&\left.4\eta^2e^{-\Gamma t}\sum_{k=1}^{\infty}\frac{(\Gamma t)^{k}}{k!}\left(\zeta(2,\eta+k)-\zeta(2,\eta)\right)\right\},
\nonumber\\
\label{eq4mom}
\end{eqnarray}
where $\zeta(a,x)$ is the Hurwitz zeta function \cite{Abramowitz1964}. The kurtosis $\kappa_{\text{a}}$ is obtained from last expression after dividing by $\langle\boldsymbol{x}^2(t)\rangle _{\text{a}}^2$ and multiplying by $d^{2}$. Clearly, the exponential prefactor in front of the sums appearing in Eq. (\ref{eq4mom}) allows us to identify immediately the asymptotic behavior of the kurtosis. Indeed, in this long-time limit, $\Gamma t\gg 1 $, the fourth moment goes with time as $\left<\boldsymbol{x}^4(t)\right>_{\text{a}}\simeq 4 d(d+2)c^4 t^2/\Gamma^2$, and the mean-squared displacement as 
$\left<\boldsymbol{x}^2\right>_{\text{a}}=2dc^2 t/\Gamma$ , thus the kurtosis has the constant value $\kappa_{\infty}\simeq d(d+2)$ as was anticipated in earlier sections.

\section{\label{AppendixActiveMotion} Hierarchy equations for active Brownian particles and run-and-tumble particles}

The hydrodynamic tensor fields mentioned in Sec. \ref{Sect-III}  are defined as follows:
\begin{subequations}\label{HydrodynamicFields}
\begin{align}
\rho(\boldsymbol{x},t)&\equiv\int d\boldsymbol{\hat{v}}~ P(\boldsymbol{x}, \boldsymbol{\hat{v}}, t),\label{Def-rho}\\
\mathbb{P}^{i}(\boldsymbol{x},t)&\equiv\int d\boldsymbol{\hat{v}} ~\boldsymbol{\hat{v}}^{i} P(\boldsymbol{x},\boldsymbol{\hat{v}},t),\label{Def_Polarization}\\
\mathbb{Q}^{ij}(\boldsymbol{x},t)&\equiv\int d\boldsymbol{\hat{v}} \left(\boldsymbol{\hat{v}}^{i}\boldsymbol{\hat{v}}^{j}-\frac{\delta^{ij}}{d}\right)P(\boldsymbol{x},\boldsymbol{\hat{v}},t),\label{Def-Nematic}\\
\mathbb{R}_{ijk}(\boldsymbol{x},t)&\equiv\int d\boldsymbol{\hat{v}} \left(\boldsymbol{\hat{v}}^{i}\boldsymbol{\hat{v}}^{j}\boldsymbol{\hat{v}}^{k}-\frac{1}{d+2}\delta^{[ij}\boldsymbol{\hat{v}}^{k]_{\text{S}}}\right)P(\boldsymbol{x},\boldsymbol{\hat{v}},t),\label{Def_Rij}
\end{align}
\end{subequations}
where $\left[\cdots\right]_{\text{S}}$ indicates that the total symmetrization of the indices must be taken into account and $P(\boldsymbol{x}, \boldsymbol{\hat{v}}, t)$ is either $P_{\text{ABP}}(\boldsymbol{x},\boldsymbol{\hat{v}}, t)$ or $P_{\text{RTP}}(\boldsymbol{x},\boldsymbol{\hat{v}}, t)$.  

After these definitions, we are going to determine the hierarchy of hydrodynamic equations for both models (\ref{ABM}) and (\ref{RT}), respectively.  The evolution equations \eqref{eqRho},\eqref{eqPol}, and  \eqref{EqQ} for $\rho(\boldsymbol{x}, t)$, $\mathbb{P}^{i}(\boldsymbol{x}, t)$, and $\mathbb{Q}_{ij}(\boldsymbol{x},t)$, respecively, are obtained by multiplying each equation (\ref{ABM}) and (\ref{RT}) with each term of the orthogonal basis, 
\begin{eqnarray}
\mathcal{B}=\left\{\boldsymbol{1}, \boldsymbol{\hat{v}}_{i}, \boldsymbol{\hat{v}}_{i}\boldsymbol{\hat{v}}_{j}-\frac{\delta_{ij}}{d},\boldsymbol{\hat{v}}^{i}\boldsymbol{\hat{v}}^{j}\boldsymbol{\hat{v}}^{k}-\frac{1}{d+2}\delta^{[ij}\boldsymbol{\hat{v}}^{k]_{\text{S}}}, \cdots\right\},\nonumber\\
\end{eqnarray}
 respectively,  and integrating out the $\boldsymbol{\hat{v}}$ degrees of freedom. In order to implement this procedure for the ABPs case we notice that it is necessary to calculate the action of $\nabla^2_{\hat{v}}$ on each element of the basis $\mathcal{B}$. For these calculations we find it useful to apply the Weingarten-Gauss (WG) structure equations, $\boldsymbol{\nabla}_{a}{\bf e}^{b}=-K_{a}^{b}{\bf n}$ and $\boldsymbol{\nabla}_{a}{\bf n}=K_{ab}{\bf e}^{b}$ valid for any  hypersurface embedded in $\mathbb{R}^{d}$ of codimension $1$, where $\boldsymbol{\nabla}_{a}$ is the covariant derivative compatible with the metric tensor $g_{ab}$,  $K_{ab}$ are the components of the extrinsic curvature tensor, $\{{\bf e}_{a}\}$ is a set of tangent vectors, and ${\bf n}$ is a normal vector at a point in the hypersurface \cite{Fal-Montiel2009}.  In the case of the hypersphere $S^{d-1}$ one has ${\bf n}=\boldsymbol{\hat{v}}$, $K_{ab}=g_{ab}$, and $\nabla^2_{\hat{v}}=\boldsymbol{\nabla}^{a}\boldsymbol{\nabla}_{a}$. For instance, the calculation  $\nabla^2_{\hat{v}}\boldsymbol{\hat{v}}$ can be carried out using the WG equations as $\nabla^2_{\hat{v}}\boldsymbol{\hat{v}}=\nabla^{c}\nabla_{c}{\bf n}=-g_{cb}g^{cb}{\bf n}=-(d-1)\boldsymbol{\hat{v}}$; similarly one can compute the Laplacian of the rest of the orthogonal basis terms. For the RTPs equations we find it useful to apply  the property that the integration $\int d\boldsymbol{\hat{v}}~\cdot$ of each element of $\mathcal{B}$ is zero as a consequence of the compactness of $S^{d-1}$. 

Next, we are going to sketch briefly the procedure used to obtain the mean-squared displacement (MSD) and the fourth moment for both ABPs and RTPs models.  In particular, using the hierarchy equations shown above we are able to provide  exact results for the MSD and the fourth moment for both models. In particular for the MSD,  one can obtain the equation  using Eq.  (\ref{MTE1})  
  \begin{eqnarray}
 \frac{d^2}{dt^2}\left<\boldsymbol{x}^2(t)\right>&+&\xi\frac{d}{d t}\left<\boldsymbol{x}^2(t)\right>=\frac{v^2_{0}}{d}\nonumber\\
 &\times&\int d^{d}\boldsymbol{x}~\boldsymbol{x}^2\left[\nabla^2\rho+v_{0}^2\partial_{i}\partial_{j}\mathbb{Q}^{ij}\right].\nonumber\\
 \end{eqnarray}
Now we perform an integration by parts, and we use the traceless property of $\mathbb{Q}^{ij}$; therefore one obtains the ordinary differential equation for $\left<\boldsymbol{x}^2(t)\right>$ 
\begin{eqnarray}
 \frac{d^2}{dt^2}\left<\boldsymbol{x}^2(t)\right>+\xi\frac{d}{d t}\left<\boldsymbol{x}^2(t)\right>=2v^2_{0}.
 \end{eqnarray}
 Under the initial conditions  $\left<\boldsymbol{x}^2(t)\right>=0$ and $d\left<\boldsymbol{x}^2(t)\right>/dt=0$ at $t=0$, one can get easily the MSD,
 \begin{eqnarray}
\langle\boldsymbol{x}^2(t)\rangle=\frac{2v_{0}^2}{\xi^2}\left[\xi t-(1-e^{-\xi t})\right],
\end{eqnarray}
where we recall that $\xi=(d-1)D_{\text{R}}$ and $\xi=\lambda$ for ABPs and RTPs models, respectively. It is noteworthy to mention that the RTPs mean-squared displacement does not depend on the dimension $d$ of the space. 

Next we are going to calculate the fourth moment of the ABPs model. The procedure implemented is similar to the one used for the MSD. Using Eq. (\ref{MTE1}) one is able to obtain the following differential equation:
\begin{eqnarray}
 \frac{d^2}{dt^2}\left<\boldsymbol{x}^4(t)\right>+\xi\frac{d}{d t}\left<\boldsymbol{x}^4(t)\right>&=&\frac{4v^2_{0}(d+2)}{d}\left<\boldsymbol{x}^2(t)\right>\nonumber\\&+&8v_{0}^2\mathcal{J}(t)\nonumber\\
 \label{eqFM}
 \end{eqnarray}
where 
 \begin{eqnarray}
\mathcal{J}(t)\equiv\int d^d\boldsymbol{x}~x_{i}x_{j}\mathbb{Q}^{ij}(\boldsymbol{x},t).
 \end{eqnarray}
This quantity can be determined using the equations of the hierarchy. In particular, using  Eq. (\ref{EqQ}) and the traceless property of $\mathbb{R}^{ijk}$ one is able to find the equation
\begin{eqnarray}
\frac{d\mathcal{J}}{d t}=-\overline{\xi}\mathcal{J}-\frac{v_{0}}{d+2}\mathcal{K}(t),
\label{eqJ}
\end{eqnarray}
where we recall the values of $\overline{\xi}$ for ABPs and RTPs models, respectively, and $\mathcal{K}(t)$ is given by
\begin{eqnarray}
\mathcal{K}(t)=\int d^{d}\boldsymbol{x}~x_{i}x_{j}\mathbb{T}^{ij}(\boldsymbol{x},t).
\end{eqnarray}
This quantity can be simplified using the structure of the tensor $\mathbb{T}^{ij}(\boldsymbol{x},t)$ and integrating by parts; thus one gets
$\mathcal{K}(t)=-2(d+2)(d-1)\mathcal{L}(t)/d$, where $\mathcal{L}(t)$ is given by
\begin{eqnarray}
\mathcal{L}(t)=\int d^{d}\boldsymbol{x}~ x_{i}\mathbb{P}^{i}(\boldsymbol{x},t).
\end{eqnarray}
Using Eq. (\ref{eqPol}) and the identity $e^{-\xi t}d(e^{\xi t}v)/dt=dv/dt+\xi v$, one is able to  find the ordinary differential equation  
$e^{-\xi t}d/dt\left(e^{\xi t} \mathcal{L}\right)=v_{0}$. Under the initial condition $\mathcal{L}(0)=0$, one can get easily the following solution:
\begin{eqnarray}
\mathcal{L}(t)=\frac{v_{0}}{\xi}\left(1-e^{-\xi t}\right).
\end{eqnarray}
Finally, using the expression for $\mathcal{L}(t)$ we get an expression for $\mathcal{K}(t)$, and  then substitute it in Eq. (\ref{eqJ}) in order to determine $\mathcal{J}(t)$. Here we also use the identity 
\begin{eqnarray}
e^{-\xi t}\frac{d}{dt}(e^{\xi t}v)=\frac{dv}{dt}+\xi v.\label{identity}
\end{eqnarray}
In particular, the equation for $\mathcal{J}(t)$ is then given by
\begin{eqnarray}
e^{-\overline{\xi} t}\frac{d}{d t}\left(e^{\overline{\xi} t}\mathcal{J}(t)\right)=\frac{2v^2_{0}(d-1)}{d\xi}\left(1-e^{-\xi t}\right).
\label{eqJ2}
\end{eqnarray}
Now we solve this equation for the initial condition $\mathcal{J}(0)=0$; then one has
\begin{eqnarray}
\mathcal{J}(t)=\frac{2(d-1)v^2_{0}}{d\overline{\xi}\xi (\xi-\overline{\xi})}\left(\xi-\overline{\xi}+\overline{\xi}e^{-\xi t}-\xi e^{-\overline{\xi}t}\right).
\end{eqnarray}
The $\mathcal{J}(t)$ solution is therefore substituted in Eq. (\ref{eqFM}) in order to determine the fourth moment $\left<\boldsymbol{x}^4\right>$. In addition, using the identity \eqref{identity}, one is able to show that Eq. (\ref{eqFM}) turns out to be 
\begin{eqnarray}
 &&e^{\xi t}\frac{d}{dt}\left[e^{\xi t}\frac{d}{dt}\left<\boldsymbol{x}^4(t)\right>\right]=\frac{8v^4_{0}(d+2)}{d\xi^2}\left[\xi t-(1-e^{-\xi t})\right]\nonumber\\&&~~~~~~~~~~~~~~~~~~~+\frac{16(d-1)v^4_{0}}{d\overline{\xi}\xi (\xi-\overline{\xi})}\left(\xi-\overline{\xi}+\overline{\xi}e^{-\xi t}-\xi e^{-\overline{\xi}t}\right).\nonumber\\
 \label{eqFM2}
 \end{eqnarray}
Under the initial conditions $\left<\boldsymbol{x}^4(t)\right>=0$ and $d\left<\boldsymbol{x}^4(t)\right>/dt=0$ at the time $t=0$, the solution of the last differential equation (\ref{eqFM2}) is given by
\begin{widetext}
\begin{multline}
\label{fourmoment}
\langle\boldsymbol{x}^4(t)\rangle=
 \frac{4v_{0}^4}{d \xi^4 \overline{\xi}^2 (\xi -\overline{\xi})^2}
\left[4(d-1)\xi^4 e^{-\overline{\xi}t}-2\overline{\xi}^2 e^{-\xi t} \left(3d\xi^2 (\xi  t+3)+(d+2)\overline{\xi}^2(\xi t+3)-2\xi\overline{\xi}(d(2\xi t+5)+\xi  t+4)\right)\right.\\
 +\left.(\xi -\overline{\xi})^2  \left(-4(d-1)\xi^2+(d+2)\overline{\xi}^2 (\xi t (\xi t-4)+6)+4(d-1)\xi \overline{\xi}(\xi t-2)\right)\right].
\end{multline}
In particular, we should take carefully the limit $\xi\to\overline{\xi}=\lambda$ for the RTPs model, which leads to the expression
\begin{multline}
\left<\boldsymbol{x}^4(t)\right>=\frac{4v_{0}^4}{d\lambda^4}\Bigl\{
6(d-4)e^{-\lambda t}+
6(d-2) t\lambda e^{-\lambda t}+
2(d-1)t^2\lambda^2e^{-\lambda t}+
\Bigl((d+2)t^2\lambda^2-12 t\lambda -6(d-4)\Bigr)
\Bigr\}.
\end{multline}
The behavior of the fourth moment for both models at the short-time regime is $\langle\boldsymbol{x}^4(t)\rangle\simeq v_{0}^4 t^4$. 
\end{widetext}

\section{\label{AppendixABM-RTM} The complete Intermediate Scattering Function two-dimensional Active Brownian Motion and Run-and-tumble motion.}
Exact expressions for the ISF for ABM and RTM can be obtained if in addition to the Fourier transform of the spatial variables $\boldsymbol{x}\rightarrow\boldsymbol{k}$, the Laplace transform of the time variable is performed, $t\rightarrow\epsilon$. These can be written in the form of continued fractions using the methods expoused in Ref. \cite{SevillaPRE2020}. For ABM we have that the partial ISF due to the active part is given by
\begin{equation}\label{ContFrac-ABM}
\widetilde{P}_\text{ABM}(\boldsymbol{k},\epsilon)=\frac{1}{2\pi}\cfrac{1}{\epsilon
	  + \cfrac{v^{2}k^{2}/2}{\epsilon+D_{R}
          + \cfrac{v^{2}k^{2}/4}{\epsilon+4D_{R}
          + \cfrac{v^{2}k^{2}/4}{\epsilon+9D_{R}+\ddots
           } } }
           },
\end{equation}
while for RTM we have
\begin{equation}\label{ContFrac-RTM}
\widetilde{P}_\text{RTM}(\boldsymbol{k},\epsilon)=\frac{1}{2\pi}\cfrac{1}{\epsilon
	  + \cfrac{v^{2}k^{2}/2}{\epsilon+\lambda
          + \cfrac{v^{2}k^{2}/4}{\epsilon+\lambda
          + \cfrac{v^{2}k^{2}/4}{\epsilon+\lambda+\ddots
           } } }
           },
\end{equation}
and both results are valid for the initial conditions considered $\delta^{(2)}(\boldsymbol{x})/2\pi$. Notice that, as has been remarked in Sec. \ref{Sect-III}, the solution \eqref{ContFrac-ABM} displays the increasing of the \emph{damping rates} $D_{R}$, $4D_{R}$, $9D_{R}$,\ldots $n^{2}D_{R}$ for the corresponding $n$-rank hydrodynamic-like field tensor of two-dimensional ABM, while the solution \eqref{ContFrac-RTM} shows the same damping rate $\lambda$ for all hydrodynamic-like field tensors of two-dimensional RTM. From these expressions, the complete ISF, i.e., the one that takes into account the effects of passive diffusion, is obtained by use of the time-shifting property of the Laplace transform, namely, by the replacement $\epsilon\rightarrow\epsilon+D_\text{T}k^{2}$.

The numerical inversion of the Laplace transform is stable and converges rather fast with the order of the approximant. The calculation has been done using \emph{Mathematica} \cite{InvLapTMathematica} and the results are shown in Fig. \ref{ISF-Comparison}.


\begin{thebibliography}{25}
\expandafter\ifx\csname natexlab\endcsname\relax\def\natexlab#1{#1}\fi
\expandafter\ifx\csname bibnamefont\endcsname\relax
  \def\bibnamefont#1{#1}\fi
\expandafter\ifx\csname bibfnamefont\endcsname\relax
  \def\bibfnamefont#1{#1}\fi
\expandafter\ifx\csname citenamefont\endcsname\relax
  \def\citenamefont#1{#1}\fi
\expandafter\ifx\csname url\endcsname\relax
  \def\url#1{\texttt{#1}}\fi
\expandafter\ifx\csname urlprefix\endcsname\relax\def\urlprefix{URL }\fi
\providecommand{\bibinfo}[2]{#2}
\providecommand{\eprint}[2][]{\url{#2}}

\bibitem[{\citenamefont{Bechinger et~al.}(2016)\citenamefont{Bechinger,
  Di~Leonardo, L\"owen, Reichhardt, Volpe, and Volpe}}]{BechingerRMP2016}
\bibinfo{author}{\bibfnamefont{C.}~\bibnamefont{Bechinger}},
  \bibinfo{author}{\bibfnamefont{R.}~\bibnamefont{Di~Leonardo}},
  \bibinfo{author}{\bibfnamefont{H.}~\bibnamefont{L\"owen}},
  \bibinfo{author}{\bibfnamefont{C.}~\bibnamefont{Reichhardt}},
  \bibinfo{author}{\bibfnamefont{G.}~\bibnamefont{Volpe}}, \bibnamefont{and}
  \bibinfo{author}{\bibfnamefont{G.}~\bibnamefont{Volpe}},
  \bibinfo{journal}{Rev. Mod. Phys.} \textbf{\bibinfo{volume}{88}},
  \bibinfo{pages}{045006} (\bibinfo{year}{2016}),
  \urlprefix\url{https://link.aps.org/doi/10.1103/RevModPhys.88.045006}.


\bibitem[{\citenamefont{Gompper et~al.}(2020)\citenamefont{Gompper, Winkler,
  Speck, Solon, Nardini, Peruani, Löwen, Golestanian, Kaupp, Alvarez
  et~al.}}]{GompperJPhysCondMatt2020}
\bibinfo{author}{\bibfnamefont{G.}~\bibnamefont{Gompper}},
  \bibinfo{author}{\bibfnamefont{R.~G.} \bibnamefont{Winkler}},
  \bibinfo{author}{\bibfnamefont{T.}~\bibnamefont{Speck}},
  \bibinfo{author}{\bibfnamefont{A.}~\bibnamefont{Solon}},
  \bibinfo{author}{\bibfnamefont{C.}~\bibnamefont{Nardini}},
  \bibinfo{author}{\bibfnamefont{F.}~\bibnamefont{Peruani}},
  \bibinfo{author}{\bibfnamefont{H.}~\bibnamefont{Löwen}},
  \bibinfo{author}{\bibfnamefont{R.}~\bibnamefont{Golestanian}},
  \bibinfo{author}{\bibfnamefont{U.~B.} \bibnamefont{Kaupp}},
  \bibinfo{author}{\bibfnamefont{L.}~\bibnamefont{Alvarez}},
  \bibnamefont{et~al.}, \bibinfo{journal}{Journal of Physics: Condensed Matter}
  \textbf{\bibinfo{volume}{32}}, \bibinfo{pages}{193001}
  (\bibinfo{year}{2020}),
  \urlprefix\url{https://doi.org/10.1088/1361-648x/ab6348}.


\bibitem[{\citenamefont{Vicsek and Zafeiris}(2012)}]{VicsekPhysRep2012}
\bibinfo{author}{\bibfnamefont{T.}~\bibnamefont{Vicsek}} \bibnamefont{and}
  \bibinfo{author}{\bibfnamefont{A.}~\bibnamefont{Zafeiris}},
  \bibinfo{journal}{Physics Reports} \textbf{\bibinfo{volume}{517}},
  \bibinfo{pages}{71 } (\bibinfo{year}{2012}), ISSN \bibinfo{issn}{0370-1573},
  \bibinfo{note}{collective motion},
  \urlprefix\url{http://www.sciencedirect.com/science/article/pii/S0370157312000968}.


\bibitem[{\citenamefont{Goldstein}(1951)}]{GoldsteinQJMAM1951}
\bibinfo{author}{\bibfnamefont{S.}~\bibnamefont{Goldstein}},
  \bibinfo{journal}{The Quarterly Journal of Mechanics and Applied Mathematics}
  \textbf{\bibinfo{volume}{4}}, \bibinfo{pages}{129} (\bibinfo{year}{1951}),
  \eprint{http://qjmam.oxfordjournals.org/content/4/2/129.full.pdf+html},
  \urlprefix\url{http://qjmam.oxfordjournals.org/content/4/2/129.abstract}.

\bibitem[{\citenamefont{Masoliver et~al.}(1993)\citenamefont{Masoliver,
  Porr{\`a}, and Weiss}}]{MasoliverPhysicaA1993}
\bibinfo{author}{\bibfnamefont{J.}~\bibnamefont{Masoliver}},
  \bibinfo{author}{\bibfnamefont{J.~M.} \bibnamefont{Porr{\`a}}},
  \bibnamefont{and} \bibinfo{author}{\bibfnamefont{G.~H.} \bibnamefont{Weiss}},
  \bibinfo{journal}{Physica A: Statistical Mechanics and its Applications}
  \textbf{\bibinfo{volume}{193}}, \bibinfo{pages}{469} (\bibinfo{year}{1993}).


\bibitem[{\citenamefont{Weiss}(2002)}]{WeissPhysicaA2002}
\bibinfo{author}{\bibfnamefont{G.~H.} \bibnamefont{Weiss}},
  \bibinfo{journal}{Physica A: Statistical Mechanics and its Applications}
  \textbf{\bibinfo{volume}{311}}, \bibinfo{pages}{381} (\bibinfo{year}{2002}).


\bibitem[{\citenamefont{Sevilla and Sandoval}(2015)}]{SevillaPRE2015}
\bibinfo{author}{\bibfnamefont{F.~J.} \bibnamefont{Sevilla}} \bibnamefont{and}
  \bibinfo{author}{\bibfnamefont{M.}~\bibnamefont{Sandoval}},
  \bibinfo{journal}{Phys. Rev. E} \textbf{\bibinfo{volume}{91}},
  \bibinfo{pages}{052150} (\bibinfo{year}{2015}),
  \urlprefix\url{http://link.aps.org/doi/10.1103/PhysRevE.91.052150}.

\bibitem[{\citenamefont{Sevilla}(2016)}]{SevillaPRE2016}
\bibinfo{author}{\bibfnamefont{F.~J.} \bibnamefont{Sevilla}},
  \bibinfo{journal}{Phys. Rev. E} \textbf{\bibinfo{volume}{94}},
  \bibinfo{pages}{062120} (\bibinfo{year}{2016}),
  \urlprefix\url{http://link.aps.org/doi/10.1103/PhysRevE.94.062120}.

\bibitem[{\citenamefont{Castro-Villarreal and Sevilla}(2018)}]{CastroPRE2018}
\bibinfo{author}{\bibfnamefont{P.}~\bibnamefont{Castro-Villarreal}}
  \bibnamefont{and} \bibinfo{author}{\bibfnamefont{F.~J.}
  \bibnamefont{Sevilla}}, \bibinfo{journal}{Phys. Rev. E}
  \textbf{\bibinfo{volume}{97}}, \bibinfo{pages}{052605}
  (\bibinfo{year}{2018}),
  \urlprefix\url{https://link.aps.org/doi/10.1103/PhysRevE.97.052605}.

\bibitem[{\citenamefont{Apaza and Sandoval}(2018)}]{ApazaSoftMatter2018J}
\bibinfo{author}{\bibfnamefont{L.}~\bibnamefont{Apaza}} \bibnamefont{and}
  \bibinfo{author}{\bibfnamefont{M.}~\bibnamefont{Sandoval}},
  \bibinfo{journal}{Soft Matter} \textbf{\bibinfo{volume}{14}},
  \bibinfo{pages}{9928} (\bibinfo{year}{2018}),
  \urlprefix\url{http://dx.doi.org/10.1039/C8SM01034J}.

\bibitem[{\citenamefont{Cates and Tailleur}(2013)}]{CatesEPL2013}
\bibinfo{author}{\bibfnamefont{M.~E.} \bibnamefont{Cates}} \bibnamefont{and}
  \bibinfo{author}{\bibfnamefont{J.}~\bibnamefont{Tailleur}},
  \bibinfo{journal}{EPL (Europhysics Letters)} \textbf{\bibinfo{volume}{101}},
  \bibinfo{pages}{20010} (\bibinfo{year}{2013}),
  \urlprefix\url{http://stacks.iop.org/0295-5075/101/i=2/a=20010}.

\bibitem[{\citenamefont{Schweitzer}(2007)}]{schweitzer2007brownian}
\bibinfo{author}{\bibfnamefont{F.}~\bibnamefont{Schweitzer}},
  \emph{\bibinfo{title}{{Brownian agents and active particles: collective
  dynamics in the natural and social sciences}}}
  (\bibinfo{publisher}{Springer}, \bibinfo{year}{2007}).

\bibitem[{\citenamefont{{Kurzthaler Christina}
  et~al.}(2016)\citenamefont{{Kurzthaler Christina}, {Leitmann Sebastian}, and
  {Franosch Thomas}}}]{KurzthalerSciRep2016}
\bibinfo{author}{\bibnamefont{{Kurzthaler Christina}}},
  \bibinfo{author}{\bibnamefont{{Leitmann Sebastian}}}, \bibnamefont{and}
  \bibinfo{author}{\bibnamefont{{Franosch Thomas}}},
  \bibinfo{journal}{Scientific Reports} \textbf{\bibinfo{volume}{6}},
  \bibinfo{pages}{36702} (\bibinfo{year}{2016}).

\bibitem[{\citenamefont{Martens et~al.}(2012)\citenamefont{Martens, Angelani,
  Di~Leonardo, and Bocquet}}]{MartensEPJE2012}
\bibinfo{author}{\bibfnamefont{K.}~\bibnamefont{Martens}},
  \bibinfo{author}{\bibfnamefont{L.}~\bibnamefont{Angelani}},
  \bibinfo{author}{\bibfnamefont{R.}~\bibnamefont{Di~Leonardo}},
  \bibnamefont{and} \bibinfo{author}{\bibfnamefont{L.}~\bibnamefont{Bocquet}},
  \bibinfo{journal}{The European Physical Journal E}
  \textbf{\bibinfo{volume}{35}}, \bibinfo{pages}{84} (\bibinfo{year}{2012}),
  ISSN \bibinfo{issn}{1292-895X},
  \urlprefix\url{https://doi.org/10.1140/epje/i2012-12084-y}.

\bibitem[{\citenamefont{Dunkel and Hänggi}(2009)}]{DunkelPhysRep2009}
\bibinfo{author}{\bibfnamefont{J.}~\bibnamefont{Dunkel}} \bibnamefont{and}
  \bibinfo{author}{\bibfnamefont{P.}~\bibnamefont{Hänggi}},
  \bibinfo{journal}{Physics Reports} \textbf{\bibinfo{volume}{471}},
  \bibinfo{pages}{1 } (\bibinfo{year}{2009}), ISSN \bibinfo{issn}{0370-1573},
  \urlprefix\url{http://www.sciencedirect.com/science/article/pii/S0370157308004171}.


\bibitem[{\citenamefont{Sevilla and Nava}(2014)}]{SevillaPRE2014}
\bibinfo{author}{\bibfnamefont{F.~J.}~\bibnamefont{Sevilla}} \bibnamefont{and}
  \bibinfo{author}{\bibfnamefont{L.~A.} \bibnamefont{Gomez~Nava}},
  \bibinfo{journal}{Physical Review E} \textbf{\bibinfo{volume}{90}},
  \bibinfo{pages}{022130} (\bibinfo{year}{2014}).

\bibitem[{\citenamefont{Porra et~al.}(1997)\citenamefont{Porra, Masoliver, and
  Weiss}}]{PorraPRE1997}
\bibinfo{author}{\bibfnamefont{J.~M.} \bibnamefont{Porra}},
  \bibinfo{author}{\bibfnamefont{J.}~\bibnamefont{Masoliver}},
  \bibnamefont{and} \bibinfo{author}{\bibfnamefont{G.~H.} \bibnamefont{Weiss}},
  \bibinfo{journal}{Physical Review E} \textbf{\bibinfo{volume}{55}},
  \bibinfo{pages}{7771} (\bibinfo{year}{1997}).

\bibitem[{\citenamefont{Kurzthaler et~al.}(2018)\citenamefont{Kurzthaler,
  Devailly, Arlt, Franosch, Poon, Martinez, and Brown}}]{KurzthalerPRL2018}
\bibinfo{author}{\bibfnamefont{C.}~\bibnamefont{Kurzthaler}},
  \bibinfo{author}{\bibfnamefont{C.}~\bibnamefont{Devailly}},
  \bibinfo{author}{\bibfnamefont{J.}~\bibnamefont{Arlt}},
  \bibinfo{author}{\bibfnamefont{T.}~\bibnamefont{Franosch}},
  \bibinfo{author}{\bibfnamefont{W.~C.~K.} \bibnamefont{Poon}},
  \bibinfo{author}{\bibfnamefont{V.~A.} \bibnamefont{Martinez}},
  \bibnamefont{and} \bibinfo{author}{\bibfnamefont{A.~T.} \bibnamefont{Brown}},
  \bibinfo{journal}{Phys. Rev. Lett.} \textbf{\bibinfo{volume}{121}},
  \bibinfo{pages}{078001} (\bibinfo{year}{2018}),
  \urlprefix\url{https://link.aps.org/doi/10.1103/PhysRevLett.121.078001}.

\bibitem[{\citenamefont{Sevilla}(2018)}]{SevillaChapter2018}
\bibinfo{author}{\bibfnamefont{F.~J.} \bibnamefont{Sevilla}},
  \emph{\bibinfo{title}{The Non-equilibrium Nature of Active Motion}}
  (\bibinfo{publisher}{Springer International Publishing},
  \bibinfo{address}{Cham}, \bibinfo{year}{2018}), pp. \bibinfo{pages}{59--86},
  ISBN \bibinfo{isbn}{978-3-319-73975-5},
  \urlprefix\url{https://doi.org/10.1007/978-3-319-73975-5_4}.

\bibitem[{\citenamefont{Sevilla et~al.}(2019)\citenamefont{Sevilla,
  Rodr\'{\i}guez, and Gomez-Solano}}]{SevillaPRE2019b}
\bibinfo{author}{\bibfnamefont{F.~J.} \bibnamefont{Sevilla}},
  \bibinfo{author}{\bibfnamefont{R.~F.} \bibnamefont{Rodr\'{\i}guez}},
  \bibnamefont{and} \bibinfo{author}{\bibfnamefont{J.~R.}
  \bibnamefont{Gomez-Solano}}, \bibinfo{journal}{Phys. Rev. E}
  \textbf{\bibinfo{volume}{100}}, \bibinfo{pages}{032123}
  (\bibinfo{year}{2019}),
  \urlprefix\url{https://link.aps.org/doi/10.1103/PhysRevE.100.032123}.

\bibitem[{\citenamefont{{Forster D.}}(1975)}]{ForsterBook}
\bibinfo{author}{\bibnamefont{{Forster D.}}},
  \emph{\bibinfo{title}{{Hydrodynamic fluctuations, broken symmetry, and
  correlation functions}}} (\bibinfo{publisher}{W. A. Benjamin, Inc., Reading,
  MA}, \bibinfo{year}{1975}),
  \urlprefix\url{https://www.osti.gov/servlets/purl/4185024}.

  \bibitem[{\citenamefont{Dulaney and Brady}(2020)}]{DulaneyPRE2020}
\bibinfo{author}{\bibfnamefont{A.~R.} \bibnamefont{Dulaney}} \bibnamefont{and}
  \bibinfo{author}{\bibfnamefont{J.~F.} \bibnamefont{Brady}},
  \bibinfo{journal}{Phys. Rev. E} \textbf{\bibinfo{volume}{101}},
  \bibinfo{pages}{052609} (\bibinfo{year}{2020}),
  \urlprefix\url{https://link.aps.org/doi/10.1103/PhysRevE.101.052609}.

\bibitem[{\citenamefont{Sevilla}(2020)}]{SevillaPRE2020}
\bibinfo{author}{\bibfnamefont{F.~J.} \bibnamefont{Sevilla}},
  \bibinfo{journal}{Phys. Rev. E} \textbf{\bibinfo{volume}{101}},
  \bibinfo{pages}{022608} (\bibinfo{year}{2020}),
  \urlprefix\url{https://link.aps.org/doi/10.1103/PhysRevE.101.022608}.

\bibitem[{\citenamefont{Kenkre and Sevilla}(2007)}]{KenkreSevilla2007}
\bibinfo{author}{\bibfnamefont{V.}~\bibnamefont{Kenkre}} \bibnamefont{and}
  \bibinfo{author}{\bibfnamefont{F.~J.} \bibnamefont{Sevilla}}, in
  \emph{\bibinfo{booktitle}{{Contributions to Mathematical Physics: a Tribute
  to Gerard G. Emch TS. Ali, KB. Sinha, eds.}}} (\bibinfo{publisher}{{Hindustan
  Book Age ncy, New Delhi}}, \bibinfo{year}{2007}), pp.
  \bibinfo{pages}{147--160}.

\bibitem[{\citenamefont{Chaikin and Lubensky}(1995)}]{chaikin_lubensky_1995}
\bibinfo{author}{\bibfnamefont{P.~M.} \bibnamefont{Chaikin}} \bibnamefont{and}
  \bibinfo{author}{\bibfnamefont{T.~C.} \bibnamefont{Lubensky}},
  \emph{\bibinfo{title}{Principles of Condensed Matter Physics}}
  (\bibinfo{publisher}{Cambridge University Press}, \bibinfo{year}{1995}).


\bibitem[{\citenamefont{Bietenholz and Giambiagi}(1995)}]{BietenholzJMathP1995}
\bibinfo{author}{\bibfnamefont{W.}~\bibnamefont{Bietenholz}} \bibnamefont{and}
  \bibinfo{author}{\bibfnamefont{J.~J.} \bibnamefont{Giambiagi}},
  \bibinfo{journal}{J. Math. Phys.} \textbf{\bibinfo{volume}{36}},
  \bibinfo{pages}{383} (\bibinfo{year}{1995}).



\bibitem[{\citenamefont{Abramowitz and Stegun}(1964)}]{Abramowitz1964}
\bibinfo{author}{\bibfnamefont{M.}~\bibnamefont{Abramowitz}} \bibnamefont{and}
  \bibinfo{author}{\bibfnamefont{I.~A.} \bibnamefont{Stegun}},
  \emph{\bibinfo{title}{Handbook of Mathematical Functions with Formulas,
  Graphs, and Mathematical Tables}} (\bibinfo{publisher}{Dover},
  \bibinfo{address}{New York}, \bibinfo{year}{1964}), \bibinfo{edition}{ninth
  dover printing, tenth gpo printing} ed., ISBN \bibinfo{isbn}{0-486-61272-4}.

\bibitem[{\citenamefont{Barrow}(1983)}]{BarrowPhilTransRoySocLond1983}
\bibinfo{author}{\bibfnamefont{J.~D.} \bibnamefont{Barrow}},
  \bibinfo{journal}{Philosophical Transactions of the Royal Society of London.
  Series A, Mathematical and Physical Sciences} \textbf{\bibinfo{volume}{310}},
  \bibinfo{pages}{337} (\bibinfo{year}{1983}).

\bibitem[{\citenamefont{Mardia}(1974)}]{Mardia74p115}
\bibinfo{author}{\bibfnamefont{K.~V.} \bibnamefont{Mardia}},
  \bibinfo{journal}{Sankhy{\=a}: The Indian Journal of Statistics, Series B}
  pp. \bibinfo{pages}{115--128} (\bibinfo{year}{1974}).

\bibitem[{\citenamefont{Hess and Klein}(1983)}]{HessKlein}
\bibinfo{author}{\bibfnamefont{W.}~\bibnamefont{Hess}} \bibnamefont{and}
  \bibinfo{author}{\bibfnamefont{R.}~\bibnamefont{Klein}},
  \bibinfo{journal}{Advances in Physics} \textbf{\bibinfo{volume}{32}},
  \bibinfo{pages}{173} (\bibinfo{year}{1983}),
  \urlprefix\url{https://doi.org/10.1080/00018738300101551}.

\bibitem[{\citenamefont{Montiel and Ros}(2009)}]{Fal-Montiel2009}
\bibinfo{author}{\bibfnamefont{S.}~\bibnamefont{Montiel}} \bibnamefont{and}
  \bibinfo{author}{\bibfnamefont{A.}~\bibnamefont{Ros}},
  \emph{\bibinfo{title}{Curves and surfaces}}, vol.~\bibinfo{volume}{69}
  (\bibinfo{publisher}{American Mathematical Soc.}, \bibinfo{year}{2009}).

\bibitem[{}]{InvLapTMathematica}
\bibinfo{author}{\bibnamefont{Wolfram Research}},
\emph{\bibinfo{title}{InverseLaplaceTransform}},
(\bibinfo{year}{2020})
\urlprefix\url{https://reference.wolfram.com/language/ref/InverseLaplaceTransform.html}.



\end{thebibliography}
\end{document}